\documentclass[a4paper,11pt]{article}

\usepackage[T1]{fontenc}
\usepackage[utf8]{inputenc}
\usepackage{geometry}
\usepackage{amsmath,amssymb,amsthm}
\usepackage{graphicx}
\usepackage{xcolor}
\usepackage{booktabs}
\usepackage{microtype}
\usepackage{multirow}
\usepackage{tabularx}
\usepackage{array}
\usepackage{enumitem}
\usepackage{float}
\usepackage[font=small]{caption}

\usepackage{fancyhdr}
\definecolor{paflBlue}{HTML}{066091}

\usepackage{hyperref}
\hypersetup{
  colorlinks=true,
  linkcolor=paflBlue,
  citecolor=paflBlue,
  urlcolor=paflBlue
}
\usepackage{doi}

\usepackage{cleveref}
\usepackage[numbers,sort&compress]{natbib}

\pdfoutput=1

\title{Physics-Attested Federated Learning:\\Securing Collaborative Anomaly Detection\\in Critical Water Infrastructure}

\author{%
  \begin{tabular}[t]{c}
    Jeff Nijsse\textsuperscript{1,*}\quad
    Shu Su\textsuperscript{2}\quad
    Benjamin Oholeguy\textsuperscript{3}\\[4pt]
    Sreenivas Sremath Tirumala\textsuperscript{4}
  \end{tabular}\\[18pt]
  \footnotesize
  \begin{tabular}[t]{@{}p{\linewidth}@{}}
    \textsuperscript{1}~Department of Software Engineering, RMIT University, Hanoi, Vietnam\\[2pt]
    \textsuperscript{2}~Department of Mathematical Sciences, Auckland University of Technology, Auckland, New Zealand\\[2pt]
    \textsuperscript{3}~Independent Researcher, New Zealand\\[2pt]
    \textsuperscript{4}~School of Science, Engineering and Technology, RMIT University, Hanoi, Vietnam\\[3pt]
    \textsuperscript{*}~Correspondence: \texttt{jeff.nijsse@rmit.edu.vn}
  \end{tabular}
}
\date{}

\begin{document}
\maketitle

\begin{abstract}
\noindent
Federated learning enables industrial operators to train shared intrusion detection models without disclosing proprietary operational telemetry. However, existing defences operate strictly in update space, leaving aggregators blind to data poisoning; model updates derived from fabricated telemetry remain indistinguishable from honest contributions. We repurpose process invariants, physical relations that always hold in normal operation, such as conservation of mass, from runtime detection heuristics into a verifiable admission requirement for federated updates, mined automatically from clean operational data. We evaluate this admission gate across two physical water testbeds (SWaT, WADI) and a distribution benchmark (BATADAL), testing seven aggregation rules against telemetry fabrication, exposure-only replay poisoning, and an invariant-aware adaptive adversary. Across three testbeds the mined invariants reject none of 25 honest shards except every cross-channel fabrication on SWaT and BATADAL, including optimised perturbations that the similarity based FoolsGold admits in full. On real telemetry, five mined invariants detect 12 of SWaT's 35 attacks, while nine invariants detect 20. With nine invariants, the admission gate removes 78\% of the targeted-attack recall lost to replay poisoning (95\% CI 57--97\%, $p < 0.001$ over 20 seeds) across five standard aggregators, and 82\% under FedAvg. To reconcile physical admission control with federated data privacy, we show invariant compliance using zero-knowledge proofs (zk-SNARKs) to allow clients to prove batch adherence without revealing operational telemetry.
\end{abstract}


\bigskip

\section{Introduction}\label{sec:intro}

Critical infrastructure networks across water, energy, and transport are increasingly targeted through operational-technology attack surfaces. In 2025 Dragos recorded 119 ransomware groups active against industrial organisations, up from 80 the previous year, having a mean dwell time of 42 days inside operational networks~\cite{dragos2026}. Water and wastewater facilities face acute operational exposure: in late 2023, the United States Cybersecurity and Infrastructure Security Agency issued an advisory confirming active exploitation of Unitronics PLCs across municipal water authorities~\cite{cisa2023}, reflecting systemic vulnerabilities documented across the sector~\cite{tuptuk2021}. Countering such intrusions requires detection mechanisms that improve as rapidly as adversaries adapt. For learning-based detectors, however, detection capability is fundamentally constrained by training data scarcity rather than model architecture~\cite{Raman2021, Koay2023}.

\begin{figure}[t]
\centering
\includegraphics[width=\textwidth]{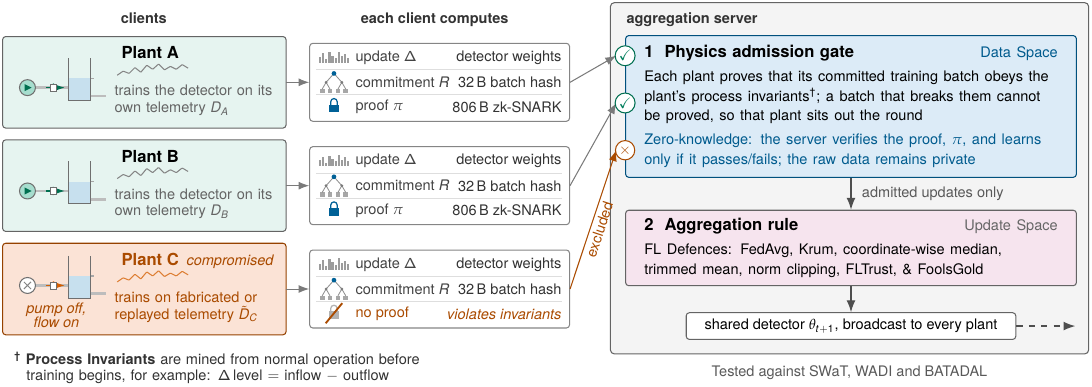}
\caption{Physics-attested federated learning. Plants train a shared intrusion detector by only exchanging model updates. We introduce a data-space admission requirement: each plant proves, in zero-knowledge, that its training batch satisfies mined process invariants (e.g., flow rates, mass balances). Batches violating these invariants cannot produce a proof and are excluded before aggregation.}
\label{fig:hero}
\end{figure}

Applying machine learning (ML) for process data analysis is standard practice, as ML-based methods can learn complex patterns and relationships within data and facilitate the automated detection of anomalies and intrusions. An anomaly detector is trained on continuous sensor streams, actuator states, set-points from supervisory control and data acquisition (SCADA) systems to learn a plant's normal operating envelope and flag physical departures from it~\cite{goh2017, kravchik2018, kravchik2022}. A principal bottleneck of this paradigm is the scarcity of localised data.

An isolated plant operates within a narrow regime, records rare documented incidents, and has virtually no labelled attack telemetry. Consequently, detectors trained at one facility generalise poorly to others. Federated learning (FL) offers an architectural remedy by training locally and transmitting only parameter updates to an aggregation server, where operators can collaboratively train a detector informed by multiple facilities while retaining proprietary telemetry on site~\cite{mcmahan2017, kairouz2021}. This formulation aligns naturally with industrial requirements, where operational data are commercially sensitive and subject to critical-infrastructure mandates such as the European Union's NIS2 Directive~\cite{eu2022}, motivating an expanding literature on federated intrusion detection in operational technology~\cite{hernandezramos2025}.

Because federated aggregation conceals local data to protect the privacy of individual participants, the coordinator cannot inspect the telemetry from which an update was derived~\cite{bagdasaryan2020, xia2023}. A single compromised operator can exploit this blind spot by training on corrupted telemetry, thereby introducing an exposure poison that teaches the shared detector to reconstruct specific attack profiles as normal operating behaviour~\cite{nguyen2020}. When the target system protects physical infrastructure, a detector conditioned to ignore an attack signature constitutes an immediate safety failure that persists throughout the adversary's dwell time. The federation thus concentrates operational risk at the exact juncture where it promised collective defence.

Existing federated defences operate exclusively in the update space. Byzantine-robust aggregation rules inspect update geometry, discounting vectors that deviate from the consensus~\cite{blanchard2017, yin2018, cao2021, fung2020}, while cryptographic input validation enforces norm bounds on hidden updates~\cite{lycklama2023, bell2023, ma2025}. Neither of these mechanisms examines the data behind the update. An update trained on entirely fabricated telemetry is indistinguishable from that computed on honest plant records. Update-space defences are known to fail against adversaries who optimise against them~\cite{fang2020, baruch2019, shejwalkar2021}, while norm bounds constrain only vector magnitude rather than directional alignment~\cite{lycklama2023}. More fundamentally, robust aggregation relies on consensus: it assumes honest updates cluster closely enough that any outlier is malicious. However, in industrial infrastructure, differing equipment, set-points, and diurnal cycles invalidate that consensus~\cite{zhou2026}. Honest updates are inherently dispersed, allowing an adversary to hide entirely within legitimate operational variance.

Industrial telemetry has an intrinsic constraint that consumer FL data lacks: it is bounded by the laws of physics. For example, mass balances dictate that tank levels change according to net flow, while hydraulic couplings require that active pumps move fluid and inactive pumps do not. These process invariants are well-established heuristics for real-time intrusion detection~\cite{adepu2016, feng2019, zhu2025, giraldo2018}. Although an adversary can construct fabricated telemetry that preserves every marginal statistic per-channel, it cannot easily satisfy multivariate physical constraints. In this paper, we convert process invariants from a detection heuristic into an \emph{admission requirement}: a mathematical condition that a client's local training batch must satisfy before its update is accepted into the round (Figure~\ref{fig:hero}).

Additionally, because these invariants are affine across measured channels, verification reduces to a set of bounded range checks that a client can prove in zero-knowledge over a committed training batch without disclosing raw telemetry. The resulting admission gate acts directly on the training data, operating before aggregation rules.

This paper investigates four primary research questions:
\begin{enumerate}[label=\textbf{RQ\arabic*:}, leftmargin=*]
  \item Do process invariants mined from normal operation separate honest operational telemetry from fabricated telemetry, without rejecting honest clients?
  \item Can an adaptive adversary train on physics-violating telemetry to craft model updates that evade robust federated aggregation?
  \item How much poisoning capability is neutralised when an adaptive adversary is forced to satisfy the invariants, and does this hold across aggregation rules?
  \item Can a participant prove batch compliance with process invariants in zero-knowledge at a cost compatible with federated rounds?
\end{enumerate}

The principal contributions of this work are fourfold:
\begin{enumerate}
  \item \textbf{A data-space admission gate}  based on automatically mined affine process invariants that rejects every synthetic fabrication breaking cross-channel relations (every cross-channel fabrication on SWaT and BATADAL) while achieving zero false rejections across 25 distinct honest client shards (10 SWaT, 10 WADI, 5 BATADAL).

  \item \textbf{An orthogonality result for data- and update-space defences.} We demonstrate across seven aggregation rules and eight attacks that four of the rules admit physics-violating fabrications at 73--100\% (including an optimised perturbation that FoolsGold admits in full), while the data gate rejects every fabrication and admits every update-space attack.

  \item \textbf{A removal-and-coverage analysis under adaptive attack.} Against an adaptive adversary that projects corrupted batches back onto the invariants, we show that the gate removes 78\% of targeted recall degradation pooled across five aggregation rules (95\% CI 57--97\%), reaching 82\% under FedAvg.

  \item \textbf{A zero-knowledge feasibility demonstration (PA-FL Lite).} We construct and benchmark a succinct zero-knowledge attestation protocol enabling clients to prove batch invariant compliance over sampled rows without disclosing operational telemetry.
\end{enumerate}

The remainder of this article is organised as follows. Section~\ref{sec:background} reviews related work in robust FL, cryptographic validation and process invariants. Section~\ref{sec:method} formalises the system model, threat framework and admission protocol. Section~\ref{sec:setup} details the experimental setup and benchmark datasets. Section~\ref{sec:results} presents the empirical results across the three operational plants. Section~\ref{sec:discussion} discusses the theoretical and practical implications, and Section~\ref{sec:conclusion} concludes the paper.

\section{Background \& Related Work}\label{sec:background}
The existing literature divides into three disconnected regimes: \textbf{(1)} robust aggregation modifies server-side vector combination heuristics, \textbf{(2)} physics-based monitoring evaluates streaming SCADA telemetry at runtime, and \textbf{(3)} cryptographic federated learning inspects encrypted gradients or proves coordinator fidelity. However, none of these paradigms verify the physical validity of the client training data.

\subsection{Byzantine-Robust Aggregation and Evasion in Federated Learning}\label{sec:background:byz-rob}
Byzantine-robust aggregation is the standard defence against poisoning. This regime addresses adversarial threats at the aggregation server. In traditional FL, distributed participants train local models on private data and transmit parameter updates to a central coordinator. Federated Averaging (FedAvg)~\cite{mcmahan2017} combines these updates via a weighted arithmetic mean. Because linear averaging grants equal or proportional trust to all submissions, a single compromised participant can arbitrarily alter the trajectory of the shared model. Byzantine-robust aggregation presents myriad rules to attempt to neutralise this vulnerability by evaluating the geometric configuration of received update vectors. Table~\ref{tab:byzantine_aggregation} summarises these representative approaches, their underlying mechanisms, and the key assumption on which their robustness depends.

\begin{table}[H]
\centering
\caption{Representative Byzantine-robust aggregation mechanisms}
\label{tab:byzantine_aggregation}
\footnotesize
{\renewcommand{\tabularxcolumn}[1]{>{\raggedright\arraybackslash}p{#1}}%
\begin{tabularx}{\linewidth}{@{}
    >{\raggedright\arraybackslash}p{2.2cm}
    >{\raggedright\arraybackslash}p{2.3cm}
    >{\raggedright\arraybackslash}X
    >{\raggedright\arraybackslash}X@{}}
\toprule
\textbf{Category} & \textbf{Method} & \textbf{Mechanism} & \textbf{Key Assumption} \\
\midrule
\multirow{2}{=}{Coordinate-wise statistics}
& Coordinate-wise median~\cite{yin2018}
& Evaluates updates coordinate-wise, selecting the median value for each parameter dimension.
& Adversarial perturbations occupy an empirical minority in every coordinate. \\
\addlinespace[0.6em]
& Trimmed mean~\cite{yin2018}
& Discards the upper and lower $\beta$-fractions along each coordinate before averaging.
& Adversarial perturbations occupy an empirical minority in every coordinate.\\
\addlinespace[0.4em]
\multirow{2}{=}{Geometric proximity}
& Krum~\cite{blanchard2017}
& Identifies the single update that minimises the sum of squared Euclidean distances to nearest neighbours.
& Benign updates form a dense spatial cluster in parameter space. \\
\addlinespace[0.4em]
& Multi-Krum~\cite{blanchard2017}
& Averages multiple low-scoring updates selected via Krum's distance metric.
& Honest updates maintain geometric coherence relative to Byzantine updates. \\
\addlinespace[0.6em]
Magnitude control
& Norm clipping~\cite{sun2019}
& Rescales update vectors whose Euclidean norm exceeds a threshold $\tau$.
& Bounding update norms limits the gradient influence of arbitrary perturbations. \\
\addlinespace[0.6em]
Trusted reference
& FLTrust~\cite{cao2021}
& Assigns client aggregation weights based on cosine similarity between each client update and a trusted reference gradient evaluated.
& Server possesses a clean, representative validation dataset. \\
\addlinespace[0.6em]
Temporal consistency
& FoolsGold~\cite{fung2020}
& Penalises client groups exhibiting sustained multi-round collinearity.
& Sybils controlled by a common adversary exhibit higher mutual similarity than honest clients. \\
\bottomrule
\end{tabularx}
}
\end{table}

Targeted poisoning strategies systematically bypass these update-space heuristics. Adversaries can inject persistent backdoors by scaling gradients to dominate the aggregate~\cite{bagdasaryan2020}, or formulate data poisoning by matching surrogate gradients~\cite{geiping2021}. When adversaries possess knowledge of the aggregation rule and benign data distribution, they can optimise perturbations to lie within the empirical coordinate variance of honest clients, evading distance-based and dimension-wise filters~\cite{fang2020, baruch2019, shejwalkar2021, xia2023}.

ML security surveys demonstrate that defences evaluated solely against static or generic attacks frequently suffer from an illusion of security; when evaluated against adaptive adversaries who possess full knowledge of the defence mechanism and optimise against it, empirical protections routinely collapse~\cite{tramer2020, carlini2019, arp2022}.

However, in these formulations, existing robust aggregation rules only evaluate the parameter vectors submitted $\Delta_i$; none inspect or constrain the underlying training telemetry $D_i$ (in Figure~\ref{fig:hero}) from which those parameters were calculated.

\subsection{Physics-Based Anomaly Detection in Industrial Control Systems}\label{sec:background:invariants}
Industrial control security relies extensively on verifying sensor-actuator consistency against the governing hydraulic, thermodynamic, and mechanical relationships~\cite{urbina2016, giraldo2018}. Process invariants formalise these relationships as algebraic constraints across monitored channels, including flow correlations across active pumps and mass balances across storage vessels. Whereas early methods manually derived invariants from process design schematics~\cite{adepu2016}, subsequent techniques automated the extraction of invariants from uncorrupted SCADA historical records using association rule mining, linear regression, and first-order predicate logic~\cite{feng2019, zhu2025}. Complementary physics-based formulations monitor sensor noise profiles~\cite{ahmed2018} or pair neural system identification with Bayesian state estimation~\cite{feng2021}, while physics-informed ML embeds the governing equations in the learned model itself~\cite{karniadakis2021}.

A central operational reality of process invariants is that their defensive capability is strictly bounded by sensor instrumentation. An audit by Cho et al.~\cite{cho2026} across the entire SWaT incident record revealed that mined invariants flagged only 13--16 of 36 physical attacks; intrusions targeting unmetered stages or shifting states within normal operational noise escape invariant violations entirely~\cite{urbina2016}.

\subsection{Cryptographic Verification in Federated Learning}\label{sec:background:crypto}
Secure aggregation protocols rely on threshold cryptography to compute parameter sums over masked updates, preventing the central coordinator from inspecting individual client models~\cite{bonawitz2017}. Because homomorphic masking conceals individual client vectors, it precludes geometric anomaly detection by the server, inadvertently shielding poisoned updates. Cryptographic input validation addresses this limitation by verifying structural properties of hidden updates in ZK. RoFL~\cite{lycklama2023} and ACORN~\cite{bell2023} employ ZK arguments to verify that encrypted updates adhere to pre-committed $\ell_2$ and $\ell_\infty$ norm bounds without exposing vector coordinates. Armadillo~\cite{ma2025} formalises privacy-utility trade-offs for such validation under malicious server threat models. In parallel, verifiable FL frameworks confirm the integrity of the aggregation, proving that the coordinator executed the aggregation arithmetic faithfully or that global models maintain specific architectural constraints, using homomorphic hashing~\cite{xu2020} or zero-knowledge proofs~\cite{wang2024, wang2025}.

Succinct zero-knowledge arguments (zk-SNARKs), notably Groth16~\cite{groth2016} combined with arithmetisation-oriented hash functions such as Poseidon~\cite{grassi2021, guo2024}, have demonstrated practical efficiency for ML inference verification and sensor telemetry commitments~\cite{chen2024, kuznetsov2026}. However, proving end-to-end model training remains computationally prohibitive because rolling backpropagation and stochastic gradient descent generate billions of arithmetic constraints per iteration~\cite{xing2025, peng2026}; proof-of-learning is cheaper, re-executing only sampled training steps, but exposes their data to the verifier~\cite{jia2021}. Available cryptographic schemes consequently verify operations on model updates or server-side arithmetic; none verify whether a participant's local training dataset represents physically valid telemetry.

\subsection{Operational Testbeds and Baseline Detectors}\label{sec:background:data}
Empirical evaluation of industrial control security relies on hardware-in-the-loop physical testbeds and high-fidelity hydraulic simulations that capture real-world cyber-physical dynamics. To this end we employ:

\textbf{SWaT (Secure Water Treatment):} A scaled operational water purification facility producing five gallons of treated water per minute in six sequential physical stages (raw water intake, chemical dosing, ultrafiltration, dechlorination, reverse osmosis, and backwash), fully instrumented with 51 physical sensors and actuators of the PLC~\cite{mathur2016, goh2016}.

\textbf{WADI (Water Distribution):} An operational water supply and distribution network testbed that captures municipal consumer distribution loops, booster pumps, and elevated storage reservoirs in 127 monitored sensor and actuator channels~\cite{ahmed2017}.

\textbf{BATADAL:} A city-scale benchmark hydraulic simulation of the C-Town distribution network (comprising seven storage tanks, 11 pumps, and 43 pipes) modelled in EPANET under realistic diurnal consumer demand~\cite{taormina2018}.

\textbf{HAI (Hardware-in-the-Loop Augmented ICS):} A multi-process testbed integrating thermal-power generation, a boiler steam turbine loop, and industrial water treatment with hardware-in-the-loop simulation~\cite{shin2021}.

Unsupervised deep learning architectures constitute the standard detection baseline across these benchmarks. In particular, 1D convolutional and recurrent reconstruction autoencoders operating on sliding windows of normalised sensor telemetry learn the plant's normal operating envelope and flag anomalies when reconstruction error exceeds a calibrated threshold~\cite{kravchik2018, kravchik2022}. Although these testbeds serve as established benchmarks for centralised intrusion detection and have been used for federated anomaly detection~\cite{huong2022}, they have not been studied with a physics-based admission check on client training data, nor under coordinated data-poisoning attacks that exploit physical process bounds.


\section{Methodology: Physics-Attested Admission}\label{sec:method}


\subsection{System Model and Threat Framework}
\begin{figure}[htb]
\centering
\includegraphics[width=1.0\textwidth]{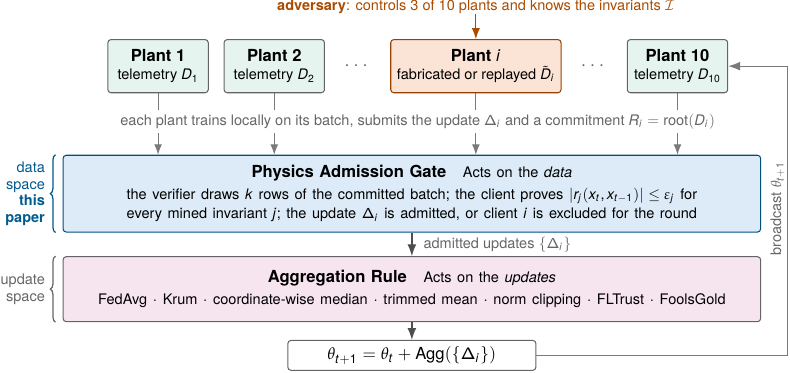}
\caption{System model and threat architecture. Ten clients collaboratively train a shared intrusion detector under synchronous federated learning with three malicious participants. The admission gate acts directly on the committed training data batch $D_i$ before the aggregation rule, providing an orthogonal defence prior to traditional update-space filtering.}
\label{fig:system}
\end{figure}

We consider a federation of $n$ industrial facilities, in this context regional water treatment plants or distribution networks,  collaboratively training a shared anomaly detection model under the orchestration of a central aggregation server (Figure~\ref{fig:system}). Honest participants operate on distinct, non-overlapping temporal shards of historical plant telemetry $D_i$, retaining proprietary operational data on site. Training progresses across synchronous communication rounds $t = 1, \dots, T$: at each round, the coordinator broadcasts the global parameter vector $\theta_t$, and each client computes local updates $\Delta_i = \theta_{t,i} - \theta_t$. Crucially, unlike conventional federated learning where client updates pass directly to server-side aggregation, our framework introduces a data-space admission gate on the local training batch prior to parameter combination.

We assume an adversary who compromises and coordinates $m$ of the $n$ participating facilities ($m = 3$ of $n = 10$ in our headline evaluations, representing a 30\% adversarial fraction, and $m = 2$ of $n = 5$ in smaller topologies). In accordance with Kerckhoffs' principle, the adversary possesses complete white-box knowledge of the federated protocol, the global parameter state $\theta_t$, its own local telemetry records, and the full set of physical invariants $\mathcal{I}$ and tolerances $\{\varepsilon_j\}$ enforced by the coordinator.

Rather than mounting an untargeted denial-of-service attack that disrupts global convergence and is readily detected via validation loss monitoring, the adversary executes targeted exposure poisoning. By injecting anomalous process states and labelling them as normal operating conditions, the adversary forces the shared reconstruction autoencoder to reconstruct these attack signatures with low error, blinding the deployed detector to subsequent real-world intrusions. We evaluate two sophistication levels: a \emph{naive attacker}, who injects raw corrupted or replayed attack telemetry directly into local batches, and a \emph{physics-aware attacker}, who explicitly projects corrupted batches onto the invariant-feasible subspace to evade the admission check while attempting to preserve poisoning efficacy. The gate inspects a client's training data, not its training. Its guarantee therefore assumes that each client computes its update from, and only from, the batch it committed. A client that commits honest data but trains on poisoned data falls outside this guarantee. We discuss this binding in Section~\ref{sec:limitations} and Appendix~\ref{sec:app:binding}.

\subsection{Process Invariants and the Admission Gate}
\label{sec:method:invariants}
\begin{figure}[t]
\centering
\includegraphics[width=\textwidth]{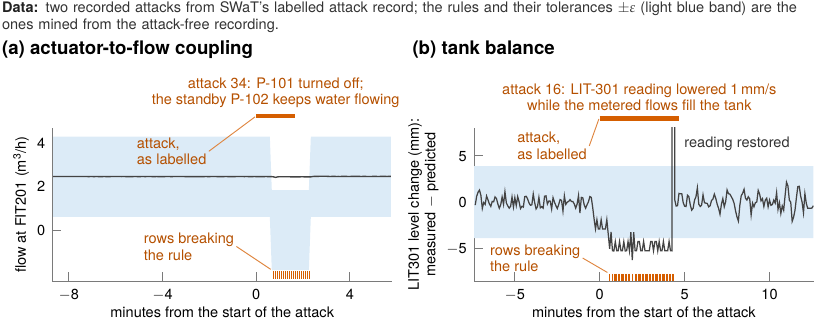}
\caption{Process invariants detecting recorded SWaT attacks: (a)~actuator-to-flow coupling during Attack~34 (P-101 off while flow persists), and (b)~tank mass balance breach during Attack~16 (sensor spoofing while filling). Shaded blue bands denote tolerances $\pm\varepsilon$; orange ticks mark rows violating the invariant ($\le 1\%$ admission threshold).}
\label{fig:invariants}
\end{figure}

Water infrastructure obeys two physical relations that an attacker who fabricates telemetry must reproduce. The first is conservation of mass in each tank: for incompressible water and a tank of constant cross-section $A_k$, the level $L_k$ changes with the net inflow,
\begin{equation*}
  A_k \frac{\mathrm{d}L_k}{\mathrm{d}t} = \sum_{\mathrm{in}} Q - \sum_{\mathrm{out}} Q,
\end{equation*}
so over a sampling interval $\Delta t$ short enough for flows to be roughly constant (5\,s here), $\Delta L_k$ is an affine function of the metered flows, with coefficients $\pm\Delta t / A_k$. The second is the operating logic of actuators: at a steady operating point a running pump or open valve passes approximately its nominal flow, while a stopped pump passes none. These relations hold only where the plant meters them (at the sensor), and transients during switching can violate them briefly. Following established physics-based monitoring~\cite{adepu2016, giraldo2018}, we write them as two families of affine invariants (Figure~\ref{fig:invariants}), whose parameters are mined from operational telemetry rather than taken from design drawings~\cite{feng2019}.

First, \textbf{Actuator-to-Flow Couplings} relate the discrete operational state of an actuator $S_j \in \{0, 1\}$ (such as a raw water intake pump or motorised chemical valve) to the continuous downstream flow measurement $F_j$:
  \begin{equation}
    r_{\mathrm{c}}(x_t) = F_j(t) - \mathbb{I}[S_j(t) = \mathrm{on}] \cdot \bar{f}_j,
    \label{eq:coupling}
  \end{equation}
  where $\bar{f}_j$ denotes the nominal steady-state volumetric flow rate and $\mathbb{I}[\cdot]$ is the indicator function. Coupling constraints are evaluated exclusively when the actuator resides in a steady operating state; telemetry rows recorded during switching transitions are marked inapplicable and excluded from residual evaluation.

Second, \textbf{Tank Mass Balances} relate differential liquid level changes across a storage or process tank $L_k$ to the net volumetric flux across pipes instrumented with flow meters:
  \begin{equation}
    r_{\mathrm{b}}(x_t, x_{t-1}) = \Delta L_k(t) - \left( \sum_{j \in \mathcal{F}_k} a_{kj} F_j(t) + c_k \right),
    \label{eq:balance}
  \end{equation}
  where $\Delta L_k(t) = L_k(t) - L_k(t - \Delta t)$ is the level change over the sampling interval, $\mathcal{F}_k$ indexes the flow meters retained by the miner, coefficients $a_{kj}$ approximate $\pm\Delta t / A_k$ for direct inflow and outflow, and offset $c_k$ absorbs unmetered losses and sensor bias. The miner regresses $\Delta L_k$ on all plant flow meters under a sparsity penalty, so the fitted balances also include flows from other process stages. These are negligible; restricting the model to the tank's inflow and outflow lowers $R^2$ by at most 0.005, and for LIT401, whose outflow is nearly constant, the inflow alone suffices.

\subsubsection{Automated Invariant Mining and Calibration}
Rather than relying on manual derivation from piping schematics, invariants are extracted automatically from an uncorrupted calibration record of normal plant operations~\cite{feng2019}. Actuator couplings are identified
by evaluating whether discrete actuator states partition continuous flow distributions into distinct, low-variance clusters, subject to minimum support and off-state ratio thresholds. Tank mass balances are identified by performing sparse linear regression of differential level changes against all metered flow rates, retaining candidate equations whose coefficient of determination satisfies $R^2 \ge 0.60$.

Because industrial instrumentation exhibits natural measurement noise, hydraulic turbulence, and minor sensor drift, invariant residuals rarely evaluate to zero under normal operation conditions.
For each mined invariant $j \in \mathcal{I}$, an operational tolerance $\varepsilon_j$ is calibrated on clean holdout telemetry:
\begin{equation}
  \varepsilon_j = 1.5 \times Q_{0.999}(|r_j|),
  \label{eq:tolerance}
\end{equation}
where $Q_{0.999}(\cdot)$ denotes the empirical 99.9th percentile of absolute residual magnitudes observed during honest operation.

\subsubsection{Batch Admission Criterion}
During federated training, a telemetry observation $x_t$ violates invariant $j$ if its absolute residual exceeds the calibrated tolerance: $|r_j(x_t)| > \varepsilon_j$. A row is classified as violating if it breaches at least one applicable invariant in $\mathcal{I}_{\mathrm{app}}(x_t)$. A candidate training batch $B$ submitted by a participant is admitted to the federated round if and only if the proportion of violating rows does not exceed the admission threshold $\alpha = 1\%$:
\begin{equation}
  \frac{1}{|B|} \sum_{x \in B} \mathbb{I}\left( \bigvee_{j \in \mathcal{I}_{\mathrm{app}}(x)} |r_j(x)| > \varepsilon_j \right) \le \alpha.
  \label{eq:admission}
\end{equation}
The $1\%$ accommodates minor transient disturbances while rejecting systematically corrupted telemetry. We evaluate two operational invariant configurations: a conservative \emph{narrow set} (5 invariants on SWaT) and an expanded \emph{wide set} (9 invariants capturing multi-stage dynamics). An attack pattern is defined as \emph{covered} by an invariant set when its constituent telemetry produces a batch violation rate exceeding $\alpha$, ensuring deterministic exclusion prior to parameter aggregation.


\subsection{Attack Formulations and Threat Vectors}
\label{sec:method:attacks}

We classify adversary behaviours across two operational representation spaces: \emph{data-space attacks} that corrupt local telemetry, $D_i$, and \emph{update-space attacks} that alter parameter vectors $\Delta_i$ directly (Table~\ref{tab:attacks}).

\begin{table}[H]
\centering
\footnotesize
\setlength{\tabcolsep}{5pt}
\caption{Classification of evaluated adversarial threat vectors across data and update spaces.}
\label{tab:attacks}
{\renewcommand{\tabularxcolumn}[1]{>{\raggedright\arraybackslash}p{#1}}%
\begin{tabularx}{\linewidth}{@{}
    >{\raggedright\arraybackslash}p{3.0cm}
    >{\raggedright\arraybackslash}p{1.8cm}
    >{\raggedright\arraybackslash}p{4.9cm}
    X@{}}
\toprule
\textbf{Attack Class / Variant} & \textbf{Locus} & \textbf{Mechanism / Preserved Property} & \textbf{Physical Relation Broken} \\
\midrule
\emph{Historical Replay}       & Data ($D_i$)   & Spliced uncorrupted physical attack segments & Only where recorded attack deviates \\
\midrule
\multicolumn{4}{@{}l}{\textbf{Statistical Fabrication}} \\
\quad \emph{Channel Roll}         & Data ($D_i$)   & Cyclic actuator shift (60 rows); preserves marginals & Actuator-to-flow couplings \\
\quad \emph{Regime Permutation}   & Data ($D_i$)   & Intra-regime actuator shuffle; preserves regime stats & Actuator-to-flow couplings \\
\quad \emph{Conservation Scaling} & Data ($D_i$)   & Flow scaling ($\times 1.3$); leaves tank levels untouched & Tank mass balances \\
\quad \emph{Regime Splicing}$^\dagger$ & Data ($D_i$) & Reordered honest segments (12 blocks); rows nominal & Segment boundary joins \\
\midrule
\emph{Optimised Perturbation}  & Data ($D_i$)   & Continuous channel shifts within $\pm 2.5\sigma$ (surrogate loss) & Both couplings and mass balances \\
\midrule
\emph{Update Baselines}        & Update ($\Delta_i$) & Sign-flip, gradient scaling, free-riding, min-max & None (data space untouched) \\
\bottomrule
\multicolumn{4}{@{}p{\linewidth}@{}}{\footnotesize $^\dagger$ Evaluated in the Separation Protocol (Section~\ref{sec:setup:protocols}) as a boundary benchmark for row-level checking.}
\end{tabularx}}

\end{table}

Data-space attacks coerce the autoencoder into learning attack patterns as nominal operations. As classified in Table~\ref{tab:attacks}, these attacks differ systematically in which physical invariant they breach. Within statistical fabrications, channel roll and regime permutation decouple actuator states from flow readings while leaving mass balances intact; conversely, conservation scaling alters flow-to-level rates and breaches mass balances while satisfying actuator couplings. Optimised perturbation violates both invariant families simultaneously, whereas historical replay breaches physical relations only during phases where the recorded attack itself contradicts instrumented physics. Regime splicing serves as a boundary case in our separation protocol, isolating single-row transition discontinuities. Conversely, update-space baselines train on uncorrupted data and modify only $\Delta_i$, establishing the reference benchmark where defence relies exclusively on server-side aggregation heuristics.

\subsubsection{Adaptive Optimisation (Physics-Aware Adversary)}
To assess the gate against an adaptive adversary~\cite{tramer2020, carlini2019}, we model a physics-aware attacker who possesses the complete invariant set $\mathcal{I}$ and tolerances $\{\varepsilon_j\}$. Given a target poisoned batch $\tilde{X}$, the adversary seeks the minimum-norm correction that renders the batch admissible:
\begin{equation}
  \min_{X} \| X - \tilde{X} \|_{\mathrm{F}}^2 \quad \text{subject to} \quad |r_j(x_t)| \le 0.95\,\varepsilon_j \quad \forall t, j \in \mathcal{I}_{\mathrm{app}}(x_t).
  \label{eq:projection}
\end{equation}
Because discrete actuator states cannot be continuously modified without immediately revealing fabrication, the adversary holds discrete actuator channels fixed and perturbs only continuous sensor readings. Rather than solving a global quadratic programme to optimality, the adversary executes an iterative active-set correction. In each pass (up to 30), it identifies active constraint violations and computes a minimum-norm step via conjugate gradients to shift continuous readings within a $0.95\,\varepsilon_j$ margin. The loop terminates as soon as the batch violation rate falls below $0.5\%$, providing a twofold safety buffer beneath the 1\% admission threshold. Batches that cannot be repaired within the 30-pass budget are submitted unamended and face gate rejection.

\subsection{Zero-Knowledge Verification Architecture (PA-FL Lite)}\label{sec:method:zk}
If clients sent raw training batches to the coordinator for inspection, they would breach data confidentiality. PA-FL Lite instead lets a client prove, without revealing telemetry, that a batch it has committed to satisfies the invariants, using Groth16 zk-SNARKs~\cite{groth2016} and the Poseidon hash~\cite{grassi2021} (protocol in Appendix~\ref{sec:app:zk}). The client commits to its batch ($N = 1{,}024$ rows) as a Poseidon Merkle root; an unpredictable challenge from the verifier selects $k = 32$ rows; and the proof shows that these rows open to the root and that at most $v_{\max}$ of them violate an invariant. A client whose proof fails is excluded from the round. Each round the client uploads only the 32-byte root and an 806-byte proof, about 0.2\% of a model update, and no telemetry. At $k = 32$ the circuit has 297{,}736 constraints and proves in 12.0\,s on commodity hardware. PA-FL Lite is a feasibility demonstration: it checks sampled rows of one committed batch and does not bind the update to that batch (Section~\ref{sec:limitations}, Appendix~\ref{sec:app:binding}).

\section{Experimental Setup}\label{sec:setup}
We outline operational testbeds, federation partitioning methodology, training parameters, and evaluation metrics. The experiment evaluates 3{,}169 runs shown in Table~\ref{tab:grid}.

\subsection{Benchmark Records and Preprocessing}\label{sec:setup:datasets}
Having established the physical testbed architectures in Section~\ref{sec:background:data}, we detail the concrete telemetry records, sampling strides, and preprocessing pipelines:
(1)~\textbf{SWaT:} We evaluate the December 2015 release (seven clean days; four days containing 36 cyber-physical attacks), downsampled to a 5-second stride with the initial 6-hour start-up transient removed. Chemical analyser (AIT) channels are excluded from detector training due to severe non-stationary electrochemical drift ($18\sigma$ baseline shifts across records; Appendix~\ref{sec:app:channels}), and discrete actuators are normalised to unit range;
(2)~\textbf{WADI:} We evaluate the October 2017 release (14 clean days; 2 attack days under 15 intrusion scenarios) downsampled to a 5-second stride, with four uninformative zero-variance alarm channels dropped and transient communication dropouts filled via linear interpolation;
(3)~\textbf{BATADAL:} We utilise the full 365-day hourly SCADA recording across 14 cyber-physical attack scenarios under realistic diurnal demand patterns; and
(4)~\textbf{HAI:} We evaluate release version 21.03 as an explicit negative control, testing non-hydraulic thermal loops where linear mass balances do not apply (Section~\ref{sec:results}).
Grouping contiguous labelled attack observations at the operational stride yields 35 distinct attack segments on SWaT (including a sustained 10-hour filtration disruption) and 14 on WADI.

\subsection{Federation Partitioning and Threat Synthesis}
\label{sec:setup:federation}

\begin{figure}[htb]
\centering
\includegraphics[width=1.0\textwidth]{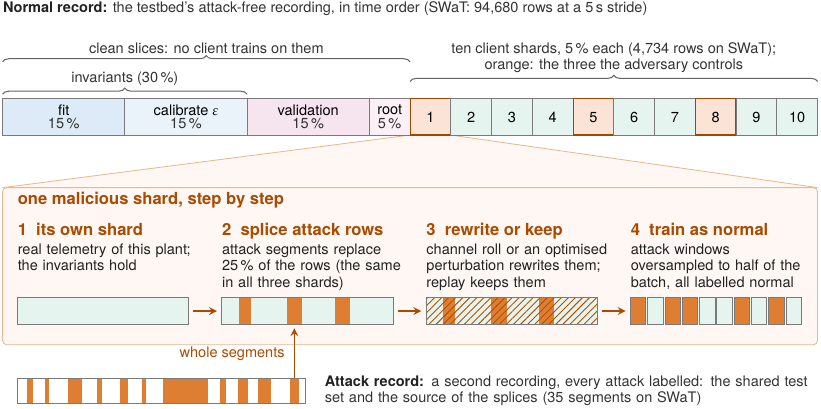}
\caption{Federation partitioning and malicious shard synthesis from a single plant record. Chronological normal telemetry is segmented into invariant mining (30\%), validation (15\%), server root (5\%), and ten disjoint client training shards (50\%). Compromised clients ($m=3$) splice attack segments across 25\% of rows and oversample attack windows to 50\% of the batch.}
\label{fig:construction}
\end{figure}

Following strict temporal separation to eliminate data leakage~\cite{arp2022}, a 10-client federation is constructed from the normal operational record of one plant, modelling a utility that operates several sites of the same design (Figure~\ref{fig:construction}). The chronological sequence is partitioned into an invariant discovery slice (30\%: half for fitting, half for tolerance calibration), a validation slice (15\%) for anomaly threshold selection, a server root slice (5\%) for FLTrust reference updates, and client training telemetry (50\%). The client partition is divided into ten disjoint temporal shards (4{,}734 rows each on SWaT at 5\,s stride). The entire attack record is held out as the global evaluation set.

In compromised federations ($m = 3$ of 10 clients on SWaT and WADI; 2 of 5 on BATADAL), malicious clients splice attack segments across 25\% of shard rows (10\% on WADI), coordinating against a single target attack set per seed. Attackers either inject statistical fabrications, optimised perturbations, invariant-projected data, or replayed historical attacks, oversampling attack windows to 50\% of local batches to ensure gradient impact.

\begin{table}[H]
\centering
\small
\setlength{\tabcolsep}{2.5pt}
\caption{Experimental grid across industrial benchmark records, detailing participant counts, aggregation rules, attack modes, federations (modes), seeds, invariant sets, and total trained models (cells).}
\label{tab:grid}
\setlength{\tabcolsep}{3pt}
\begin{tabular}{@{}l c c c c c l r@{}}
\toprule
Record & \begin{tabular}[b]{@{}c@{}}Clients\\{}(malicious)\end{tabular} & Rules & Attacks & Modes & Seeds & Invariant sets & Cells \\
\midrule
SWaT & 10 (3) & 7 & 8 & 5 & 20 & narrow (5), wide (9) & 3007 \\
WADI & 10 (3) & 3 & 2 & 5 & 3 & default (7) & 90 \\
BATADAL & 5 (2) & 3 & 2 & 4 & 3 & mined (10) & 72 \\
\bottomrule
\end{tabular}

\end{table}

\subsection{Local Detector Architecture and Training Setup}
\label{sec:setup:training}

Following standard ICS anomaly detection benchmarks~\cite{kravchik2018, kravchik2022}, each facility trains an unsupervised deep reconstruction autoencoder\footnote{We use a standard architecture without tuning, and evaluate the gate by admission and the detector by the damage that admitted poison causes.} over sliding temporal windows of normalised telemetry ($W = 10$). The network compresses each flattened window through a 128-unit dense layer with ReLU activations into a 24-dimensional latent bottleneck, which a symmetric decoder reconstructs under mean squared error. Observations whose reconstruction error exceeds the 99.5th percentile of clean validation loss ($\tau_{\mathrm{det}}$) trigger intrusion alarms.

Federated training executes for 25 rounds (2 local epochs per round, Adam with $\eta = 10^{-3}$, mini-batches of $B = 128$ windows). On commodity hardware, a full 25-round run completes in $~17$ seconds, enabling exhaustive evaluation across the 3{,}169-run experimental grid (Table~\ref{tab:grid}). Each seeded run redraws the malicious clients, the target attack set, the fabrication, and the weight initialisation keeping the temporal shards fixed. Every removal ratio is measured on SWaT across 20 seeds, and WADI and BATADAL across 3.\footnote{WADI and BATADAL use 3 seeds each: on WADI the naive attack never costs a point of targeted recall, and on BATADAL the gate excludes no attacker, so more seeds cannot add removal evidence. Both records serve separation and admission.}

\subsection{Baseline Aggregation Rules}
\label{sec:setup:aggregation}

Because the admission gate operates in data space prior to local gradient computation, it is designed to complement server-side aggregation. To test orthogonality across the defensive paradigms introduced in Section~\ref{sec:background:byz-rob}, we benchmark the gate alongside seven representative aggregation rules:
(1)~unfiltered baseline: FedAvg;
(2--3)~coordinate-wise estimators: Median and Trimmed Mean (trimming fraction $\beta = 20\%$);
(4--5)~geometric filters: Norm Clipping (threshold $\tau$) and Krum (configured with Byzantine allowance $f = m = 3$); and
(6--7)~directional and historical scoring: FLTrust (evaluated against the 5\% server root partition, see Section~\ref{sec:setup:federation}) and FoolsGold (tracked across all 25 rounds).

\subsection{Federated Execution Modes and Evaluation Metrics}
\label{sec:setup:metrics}

To isolate defensive mechanics, experiments evaluate five federated operational modes:
(1)~\textbf{Clean}, where all $n = 10$ participants train honestly;
(2)~\textbf{Honest-Only}, where the $n - m = 7$ honest clients train in isolation, establishing the reference baseline that isolates poisoning damage from changes in collective participant capacity;
(3)~\textbf{Naive Attack}, where 7 honest and 3 malicious clients aggregate without data-space admission checks;
(4)~\textbf{Physics-Aware Attack}, where projected malicious updates are aggregated unconditionally; and
(5)~\textbf{Gated}, the deployed system where batches breaching Equation~\eqref{eq:admission} are excluded prior to aggregation ($w_i = 0$).

The threat model is strictly targeted: the adversary trains the shared autoencoder to reconstruct specific attack patterns as nominal operations, leaving benign and untargeted operations unperturbed. Because normal operations constitute  $88$\% of the test record (attacks $<12\%$), aggregate metrics averaged throughout the test record barely register severe disruptions confined to brief attack segments---a well-documented pitfall in security evaluation~\cite{arp2022, cho2026}. Blinding a detector to three targeted attacks alters global F1 by $<0.01$ points~\cite{bagdasaryan2020}, and under FedAvg the four data-space attacks lower targeted recall by 6.0 to 8.6 points (11.7 points in our 20-seed replay experiment) while global F1 does not fall at all (Table~\ref{tab:detector}). A defender relying on F1 would judge the deployed model uncompromised.

We therefore define \emph{Targeted-Attack Recall} ($R_{\text{targ}}$) as the primary metric, measuring detection within the coordinated target attack set $\mathcal{A}_{\text{targ}}$:
\begin{equation}
  R_{\text{targ}} = \frac{\sum_{t \in \mathcal{A}_{\text{targ}}} y_t \cdot \hat{y}_t}{\sum_{t \in \mathcal{A}_{\text{targ}}} y_t},
  \label{eq:targ_recall}
\end{equation}
where $y_t \in \{0, 1\}$ is the ground-truth label and $\hat{y}_t \in \{0, 1\}$ indicates whether reconstruction error exceeds threshold $\tau_{\mathrm{det}}$.

The \emph{Damage-Weighted Removal Ratio} quantifies the proportion of naive poisoning damage removed by the admission gate relative to the Honest-Only reference baseline:
\begin{equation}
  \text{Removal} = 1 - \frac{R_{\text{ref}} - R_{\text{gated}}}{R_{\text{ref}} - R_{\text{naive}}},
  \label{eq:removal}
\end{equation}
where $R_{\text{ref}}$, $R_{\text{naive}}$, and $R_{\text{gated}}$ denote targeted recall under the honest reference baseline, naive attack, and gated admission, respectively. Removal is evaluated over informative seeds where naive attack damage is measurable ($R_{\text{ref}} - R_{\text{naive}} \ge 1.0\%$). Evaluation metrics thus decouple into two distinct families: gate-level metrics evaluate data admission (honest shards rejected and malicious batches excluded; Table~\ref{tab:admission}), while detector-level metrics measure the operational damage that admitted poison causes to the federated detector (targeted and untargeted recall, global F1, and AUC-PR; Appendix Table~\ref{tab:detector}).

\subsection{Verification Protocols and Reproducibility}
\label{sec:setup:protocols}

Invariant performance is verified prior to federated training via two protocols (detailed in Appendix~\ref{sec:app:protocols}): (1) a \emph{Separation Protocol} evaluating 200 random 1{,}000-row batches per record across clean and fabricated telemetry to verify zero false rejections; and (2) a \emph{Coverage Audit} scoring all 35 SWaT and 14 WADI attacks individually against narrow and wide invariant sets.

The federated simulation, invariant miner and analysis scripts are implemented in Python 3.13 with PyTorch 2.14, and the circuit in Circom 2.1.9 with SnarkJS 0.7.4; all runs use an Apple M3 Pro (36\,GB RAM, CPU only), with hyperparameters fixed a priori and no search. For more information, refer to the Data Availability Statement.

\section{Results}\label{sec:results}

\subsection{RQ1: Separating Honest from Fabricated Telemetry}\label{sec:results:rq1}

\begin{figure}[htb]
\centering
\includegraphics[width=1.0\textwidth]{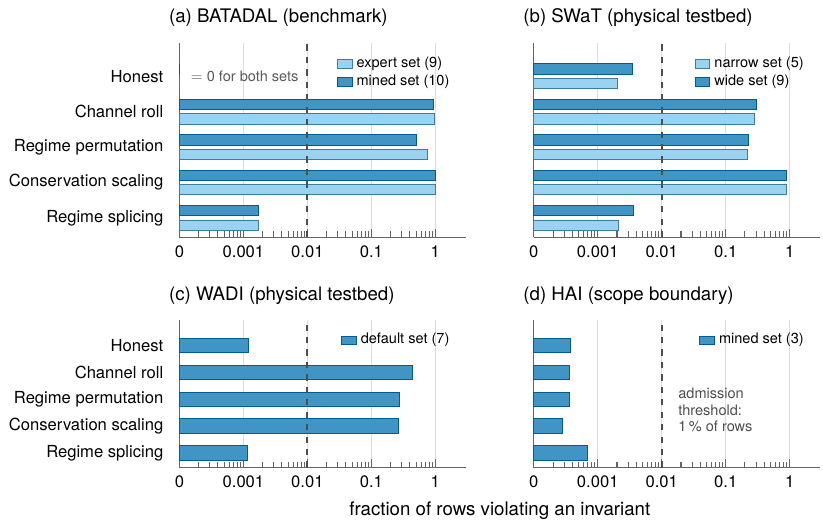}
\caption{Separation of honest from fabricated telemetry by process invariants. Mean fraction of rows violating at least one invariant over 200 batches of 1{,}000 rows (log scale; zero plotted at $10^{-4}$). The vertical dashed line marks the 1\% admission threshold $\alpha$. (a)~BATADAL under expert (9) and mined (10) invariant sets. (b)~SWaT under narrow (5) and wide (9) mined invariant sets. (c)~WADI under the default mined set (7 couplings). (d)~HAI is a scope boundary since the miner finds no couplings and only three weak balances, so every batch is admitted.}
\label{fig:c1-separation}
\end{figure}

We first ask whether process invariants mined from normal operation can separate honest operational telemetry from fabricated batches without rejecting honest participants. Following our Separation Protocol (Section~\ref{sec:setup:protocols}), we evaluate 200 randomly placed batches of $N = 1{,}000$ consecutive observations per benchmark under the 1\% admission threshold $\alpha = 0.01$ (Figure~\ref{fig:c1-separation}).

Across the three benchmarks (SWaT, WADI, and BATADAL), honest data is not rejected ($\mathrm{FRR} = 0.000$). Across 25 distinct honest client shards (10 SWaT, 10 WADI, 5 BATADAL), the highest observed single-shard row violation rate is $0.51\%$ on SWaT (under wide invariants), which sits safely below the 1\% admission gate. In contrast, when data statistics are fabricated, they violate physical constraints:
(1)~\emph{Channel Roll} desynchronises control commands from hydraulic responses, violating invariants in $28.1\%$ to $96.8\%$ of batch rows;
(2)~\emph{Within-Regime Permutation} reassigns actuator states, producing violation rates of $22.2\%$ to $75.6\%$; and
(3)~\emph{Conservation Scaling} disrupts mass and actuator-flow balances, generating violation rates between $26.9\%$ to $99.0\%$.
These are batch means; per batch, WADI, which has no mass balances, is the exception: its check admits 3\% of channel-roll and 23.5\% of conservation-scaling batches, against none on SWaT and BATADAL.

Two boundaries emerge from Figure~\ref{fig:c1-separation}. First, \emph{Regime Splicing} produces an empirical violation rate of only $0.12\%$ to $0.36\%$. Because spliced sequences consist of genuine historical telemetry segments, physical laws are satisfied internally throughout each contiguous segment, triggering violations exclusively across splice boundaries. Spliced fabrications thus pass row-level percentage checks. Second, \textit{HAI}, a boiler and steam-turbine testbed, has almost no actuator switching and no tank-style mass balance in its training record, so the miner finds no couplings and only three weak balances. HAI marks our work's boundary and confirms that the framework applies strictly to hydraulic and mass-conserving physical processes.

In sum, yes, process invariants mined from normal operation can separate honest from fabricated telemetry within the coverage of the invariants.


\subsection{RQ2: Evasion of Robust Aggregation and Data-Space Admission}\label{sec:results:rq2}
\begin{figure}[htb]
\centering
\includegraphics[width=1.0\textwidth]{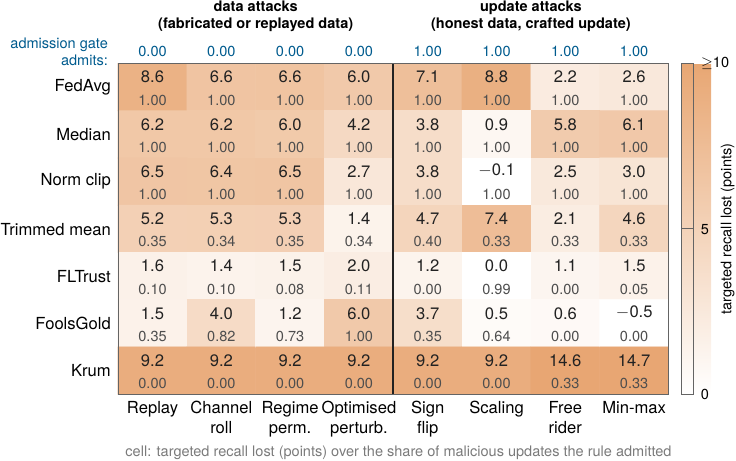}
\caption{Blind spots of the two defences on SWaT (wide invariant set, no gate 3 seeds). Rows are the aggregation rules; columns are the four data-space attacks on the left and the four update-space attacks on the right. Each cell gives the targeted-attack recall lost in points (upper number and colour) and, below it, the share of malicious updates the rule admitted.  The top row is the share of malicious batches the admission gate admits: 0.00 for every data attack, whose batches break the invariants, and 1.00 for every update attack, whose clients train on honest data and tamper only with the update. Each defence is blind to one half of the grid.}
\label{fig:complementarity}
\end{figure}

We next ask whether updates trained on fabricated or replayed telemetry evade update-space robust aggregation, and whether data-space admission closes that blind spot. The damage is measured based on targeted-attack recall as in Section~\ref{sec:setup:metrics}. Figure~\ref{fig:complementarity} reports the targeted recall loss and client acceptance rates across seven aggregation rules under naive data-space attacks (left block) and update-space attacks (right block) in the absence and presence of the admission gate.

Across all four data-space attacks (replay, channel roll, permutation, and optimised perturbation), update-space aggregation rules fail to prevent model poisoning.

\textbf{Coordinate estimators and geometric filters:} FedAvg, coordinate-wise median and norm clipping admit 100\% of the poisoned updates and lose 6.0 to 8.6, 4.2 to 6.2 and 2.7 to 6.5 points respectively across the data-space attacks. Norm clipping rescales the update but keeps its direction, so it behaves like FedAvg. Trimmed mean discards the extreme 20\% of each coordinate and admits only 34\% of the malicious updates, yet still loses 5.2 to 5.3 points under replay, channel roll and regime permutation (1.4 under optimised perturbation). Krum admits none of them and still shows a uniform 9.2-point drop on every attack: that drop is selection variance, since Krum forwards a single client's update, and not poisoning.

\textbf{Directional and similarity-based defences:} FLTrust admits 8\% to 11\% of the poisoned updates by comparing them with a clean server root and holds the loss to 1.4 to 2.0 points. FoolsGold, which penalises updates that stay collinear across rounds, admits 100\% of the optimised perturbation updates and loses 6.0 points on them, and admits 82\% under channel roll (4.0 points).

\textbf{Structural division:} Figure~\ref{fig:complementarity} illustrates a structural division between data-space admission and update-space aggregation rules\label{sec:results:complementarity}. When malicious participants poison training data (the four data-space attacks on the left), update-space defences are ineffective, while the admission gate rejects 100\% of these batches ($w_i = 0$), preventing corrupted gradients from entering aggregation. Conversely, when clients train on honest telemetry and manipulate only the parameter update (the four update-space attacks on the right: sign flip, scaling, free rider, min-max), the admission check admits every batch ($w_i = 1.0$), and the aggregation rule is the sole defence. Here performance is uneven: FLTrust holds every update attack to $\le 1.5$ points and FoolsGold resists all but sign flip, whereas FedAvg loses 7.1 and 8.8 points to sign flip and scaling, median and trimmed mean lose 4 to 7 points, and Krum is degraded heavily by free rider and min-max (14.6 and 14.7 points). The choice of aggregation rule thus governs update-space resilience, while data-space admission neutralises data poisoning.

In summary, to answer RQ2: Do
updates trained on fabricated telemetry evade update-space robust aggregation? Four of the seven aggregation rules admit 73--100\% of the updates trained on poisoned telemetry, and FoolsGold admits the optimised perturbation in full. The gate rejects every data-space attack and admits every update-space attack, so the two defences cover complementary halves of the threat and are needed together.

\subsection{RQ3: Damage Removal under an Adaptive Adversary}\label{sec:results:rq3}
Next, we investigate how much targeted damage admission removes when an adaptive adversary knows the invariants and projects its poisoned data onto them, and how this depends on invariant coverage.

\begin{figure}[htb]
\centering
\includegraphics[width=1.0\textwidth]{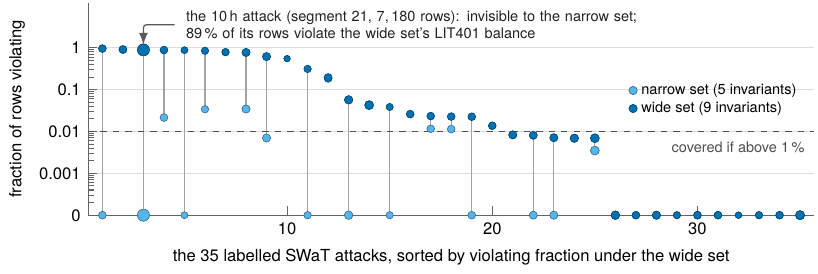}
\caption{SWaT attacks identified by invariant set. The fraction of rows violating at least one invariant for each of the 35 labelled SWaT attacks under narrow (5 invariants) and wide (9 invariants) sets, sorted by the wide-set value (log scale; zero plotted at $10^{-4}$). Stems join each attack's narrow and wide values; larger markers denote longer attacks; the dashed line is the 1\% admission threshold: an attack to its right is covered.}
\label{fig:coverage}
\end{figure}

\textbf{Narrow vs.\ wide attack coverage.}\label{sec:results:narrow-wide}
Because an admission gate operates by evaluating algebraic invariant residuals, it can exclude only those poisoning attacks whose underlying telemetry contradicts instrumented physical relations. We quantify this relationship across the 35 labelled cyber-physical attacks on SWaT and 14 on WADI (Figure~\ref{fig:coverage} and Table~\ref{tab:coverage}).

\begin{table}[H]%
\centering
\small
\caption{Coverage of labelled attacks across invariant sets on SWaT and WADI. An attack segment is covered when its data violation rate strictly exceeds the 1\% threshold. Honest row violations are audited over the first 60{,}000 observations following the discovery slice.}
\label{tab:coverage}
\setlength{\tabcolsep}{3pt}
\begin{tabular}{@{}l c c c c r c@{}}
\toprule
Record, setting & Rules & \begin{tabular}[b]{@{}c@{}}Couplings /\\balances\end{tabular} & \begin{tabular}[b]{@{}c@{}}Honest rows\\violating\end{tabular} & \begin{tabular}[b]{@{}c@{}}Attacks\\covered\end{tabular} & \begin{tabular}[b]{@{}c@{}}Attack rows\\covered\end{tabular} & Share \\
\midrule
SWaT, narrow & 5 & 3 / 2 & 0.21\,\% & 12 / 35 & 1,137 / 10,931 & 10\,\% \\
SWaT, wide & 9 & 6 / 3 & 0.36\,\% & 20 / 35 & 8,991 / 10,931 & 82\,\% \\
WADI, default & 7 & 7 / 0 & 0.14\,\% & 3 / 14 & 395 / 1,996 & 20\,\% \\
\bottomrule
\end{tabular}

\end{table}

As shown in Figure~\ref{fig:coverage} and Table~\ref{tab:coverage}, invariant coverage functions as an operational dial. Under conservative miner defaults ($R^2 \ge 0.60$, support $0.020$, coupling off-ratio 0.05), the narrow set keeps only actuator couplings and single-tank balances, and cannot see attacks that move water between stages such as the 10-hour ultrafiltration backwash drain (\textit{Segment 21}), that violates none of its invariants and passes the check unseen. Relaxing the thresholds ($R^2 \ge 0.40$, support $0.005$, coupling off-ratio 0.10) a setting we introduced after the narrow set missed this attack, incorporates multi-stage mass balances P1 through P4: the same attack now violates the LIT401 balance in 89\% of its rows while increasing the honest violation rate to only 0.36\%, safely below the 1\% threshold. On the WADI distribution network, the miner discovers 7 actuator couplings but zero reliable mass balances due to unmetered consumer demand loops, covering 3 of 14 attacks (19.8\% of rows) and demonstrating that physical coverage is fundamentally bounded by metering capability.

\begin{figure}[htb]
\centering
\includegraphics[width=\textwidth]{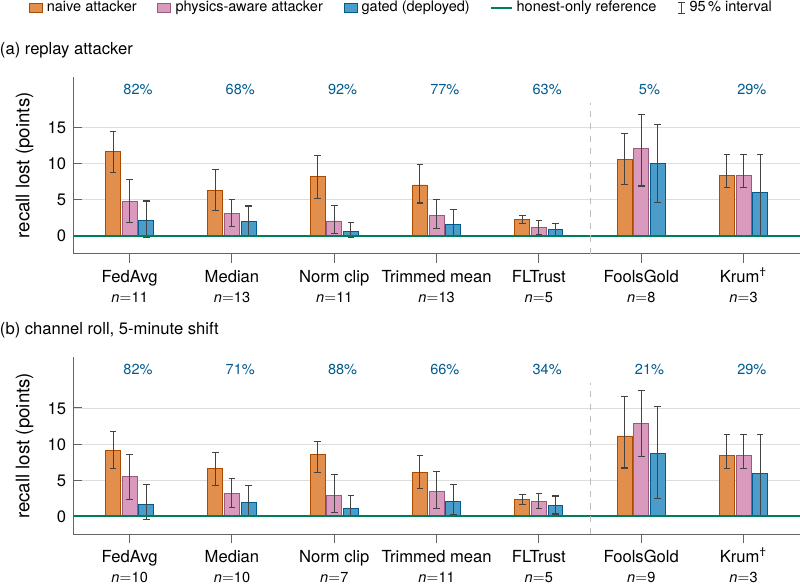}
\caption{Poison damage removal across seven aggregation rules on SWaT under the wide invariant set (10 clients, 3 malicious, 20 seeds). Targeted-attack recall lost is evaluated across four federated execution modes: \emph{Honest-Only} (unpoisoned reference at $0$), \emph{Naive} (unfiltered poisoned training), \emph{Physics-Aware} (adversary projects telemetry onto invariants), and \emph{Gated} (deployed server rejects non-compliant batches). Subplots show (a)~historical attack replay and (b)~channel roll (5-minute shift). Bars are means over the $n$ informative seeds (naive damage $\geq 1$ point) with 95\% bootstrap intervals; percentages give the share of naive damage removed by the gate. $^\dagger$Krum admits no malicious update, so its damage reflects which honest update it selects.}
\label{fig:removal}
\end{figure}

\textbf{Poison damage removal across aggregation rules.}\label{sec:results:removal}
Having established that robust aggregation cannot prevent data-space poisoning, we examine whether process physics can neutralise this threat across all seven aggregation rules (Figure~\ref{fig:removal} and Appendix Table~\ref{tab:removal}). We evaluate two distinct defensive barriers relative to the unpoisoned reference: first, whether forcing an adaptive adversary to project poisoned telemetry onto physical invariants diminishes attack potency, and second, whether filtering non-compliant batches at the admission gate restores detection performance.

\begin{table}[H]
\centering
\small
\caption{Batch admission and client exclusion rates across industrial benchmark records before and after adaptive projection under the 1\% gate. \emph{Malicious excluded} is the mean per run. Honest shards are counted once each.}
\label{tab:admission}
\setlength{\tabcolsep}{4pt}
\setlength{\tabcolsep}{3pt}
\begin{tabular}{@{}l c c c c c c@{}}
\toprule
 & \multicolumn{3}{c}{malicious batches admitted} & & \multicolumn{2}{c}{honest shards} \\
\cmidrule(lr){2-4}\cmidrule(lr){6-7}
Record (rules) & fabricated & \begin{tabular}[b]{@{}c@{}}replay,\\{}before\end{tabular} & \begin{tabular}[b]{@{}c@{}}replay,\\{}after proj.\end{tabular} & \begin{tabular}[b]{@{}c@{}}malicious\\{}excluded\end{tabular} & rejected & \begin{tabular}[b]{@{}c@{}}worst\\{}violating\end{tabular} \\
\midrule
BATADAL (10) & 0 / 6 & 0 / 6 & 6 / 6 & 0.0 of 2 & 0 / 5 & 0.00\,\% \\
SWaT narrow (5) & 0 / 60 & 24 / 60 & 51 / 60 & 0.5 of 3 & 0 / 10 & 0.32\,\% \\
SWaT wide (9) & 0 / 60 & 3 / 60 & 38 / 60 & 1.1 of 3 & 0 / 10 & 0.51\,\% \\
WADI (7) & 0 / 9 & 1 / 9 & 3 / 9 & 2.0 of 3 & 0 / 10 & 0.13\,\% \\
\bottomrule
\end{tabular}

\end{table}

Forcing an adaptive adversary to satisfy the physical invariants already costs it a substantial share of its poisoning damage. Projection alone removes 50\% to 75\% of replay damage under FedAvg, median, trimmed mean and norm clipping, yet leaves 63\% of projected batches admissible to the gate (Table~\ref{tab:admission}). The threshold $\alpha = 1\%$ is the tightest setting tested in the sweep that rejects no honest shard (Appendix~\ref{sec:app:alpha}). Because discrete actuator states remain fixed (e.g., on/off), satisfying the invariants forces sensor readings into unnatural configurations\footnote{In one projected batch, flow meter FIT401 reads $-62$ m$^3$/h.}. This distortion weakens the poison but does not prevent the model from learning the attack.

The admission gate removes most of the remaining damage. Under FedAvg, it removes 82\% of naive replay damage (95\% CI 61--102\%), from an 11.7-point drop in targeted recall, and 82\% of channel-roll damage (57--105\%). Median, trimmed mean and norm clipping retain 66\% to 92\% of the benefit across both attackers, each significant at $p \le 0.016$ (exact one-sided Wilcoxon test; Table~\ref{tab:removal}).\footnote{Interval bounds above $100\%$ arise when gated runs slightly outperform the unpoisoned reference ($R_{\text{gated}} > R_{\text{ref}}$ in Equation~\eqref{eq:removal}) within seed variance.} Pooled over the five rules with the seed as the unit, the gate removes 78\% of the damage (95\% CI 57--97\%, $p < 0.001$), significantly more than projection alone ($p = 0.008$). The variance of removal across seeds is structural, following the number of attackers the gate excludes. Removal is 19--40\% in seeds where no attacker is excluded, 75--86\% where one is, and 100\% where all three are. FLTrust, which already admits few malicious updates, gains less: 63\% of replay damage and 34\% of channel-roll damage, the latter not significant. FoolsGold shows no reliable removal (intervals span zero), and Krum admits no malicious update, so its residual damage reflects which honest update it selects rather than the poison.

\begin{figure}[htb]
\centering
\includegraphics[width=1.0\textwidth]{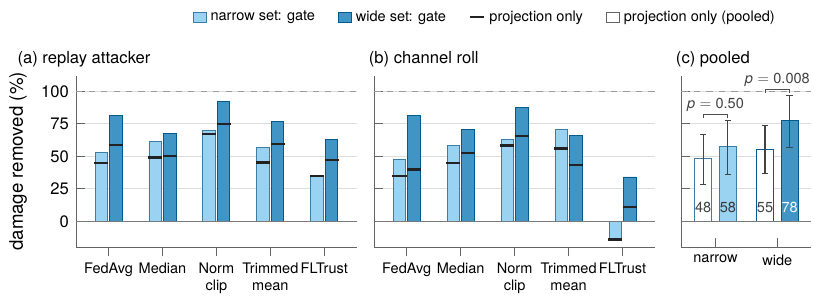}
\caption{(a, b) The percentage of targeted damage removed by the deployed gate under narrow versus wide invariant sets on SWaT. The solid line marks removal by projection alone, so the gap above is the gate's contribution. (c)~All five rules combined: removal by projection alone (open bars) and by the gate (filled bars), with 95\% bootstrap intervals.}
\label{fig:dial}
\end{figure}

\textbf{Effect of invariant coverage on damage removal.} To confirm that damage removal is governed by physical invariant coverage rather than extraneous artefacts, Figure~\ref{fig:dial} contrasts gate removal under the narrow set (12/35 attacks, 10\% rows) against the wide set (20/35 attacks, 82\% rows) across identical seeds. Under the narrow set, the admission gate admits 85\% of projected batches, excluding an average of only $0.5$ of 3 malicious clients per run (attackers are excluded in 3 of 20 seeds); its removal of 53\% to 70\% of replay damage is almost entirely the effect of projection, and the gate adds no significant removal beyond it (pooled 48\% to 58\%, $p = 0.50$). Under the wide set, the additional multi-stage mass balances reduce batch admission to 63\%, excluding an average of $1.1$ of 3 malicious clients per run, and the gate raises removal from 55\% under projection to 78\% (pooled, $p = 0.008$; 66\% to 92\% per rule across both attackers). Wider invariant coverage leaves more projected batches unable to comply, and it is these exclusions that let the gate remove damage beyond what projection alone achieves.

In summary, projection alone already costs the adversary about half its damage (55\% pooled). Under the wide invariant set the gate removes most of the rest by excluding the batches that projection cannot bring under $\alpha$: 78\% pooled (57--97\%), significantly more than projection ($p = 0.008$), and 66--92\% for FedAvg, median, trimmed mean and norm clipping. Under the narrow set it adds no significant removal ($p = 0.50$). Invariant coverage decides how much the gate adds.

\subsection{RQ4: Zero-Knowledge Verification Feasibility}\label{sec:results:rq4}

Finally, we evaluate whether a participant can prove batch compliance with process invariants in zero-knowledge at a cost compatible with federated rounds, without disclosing proprietary operational telemetry.

\textbf{Cryptographic benchmarks and circuit overhead.}
We evaluate the PA-FL Lite protocol instantiated with Groth16 zk-SNARKs over the BN128 elliptic curve on an Apple M3 Pro workstation (Table~\ref{tab:zk-headline}). PA-FL Lite achieves practical operational efficiency at the $k = 32$ operating point (with complete $k = 16$ benchmarks detailed in Appendix Table~\ref{tab:zk-full}). The Circom circuit compiles to $297{,}736$ R1CS constraints ($9{,}304$ constraints per sample, dominated by chunked Poseidon hashing and Merkle opening paths, compared to $148{,}888$ constraints at $k = 16$). On the local plant client, generating the attestation proof $\pi_i$ requires $12.0\,$s and $3.8\,$GB peak RAM ($6.55\,$s and $2.4\,$GB at $k = 16$).

\begin{table}[H]
\centering
\small
\caption{PA-FL Lite at the operating point $k = 32$ on wide SWaT with $N = 1{,}024$ rows per batch (Groth16 over BN128, Apple M3 Pro): the cost of one attestation and the sampling guarantee it buys. The full benchmark, including
$k = 16$, is Appendix Table~\ref{tab:zk-full}.}
\label{tab:zk-headline}
\setlength{\tabcolsep}{6pt}
\begin{tabular}{@{}l r@{}}
\toprule
\multicolumn{2}{@{}l}{\textbf{PA-FL Lite at $k = 32$ sampled rows}} \\
\midrule
Circuit size (R1CS constraints) & 297{,}736 \\
Proof generation on the plant (peak RAM) & 12.0\,s (3.8\,GB) \\
Proof verification at the server & 290\,ms \\
Proof size $\pi_i$ & 806\,B \\
Honest batch accepted, $P(v \le 1)$, 19{,}992 draws & 98.5\,\% \\
Replayed attack batch detected, $P(v > 1)$ & 99.8\,\% \\
Channel-roll batch detected, $P(v > 1)$ & 99.8\,\% \\
\bottomrule
\end{tabular}

\end{table}

On the coordinator side, proof verification executes in just $290\,$ms, while the submitted Groth16 proof $\pi_i$ consumes only $806$ bytes of communication bandwidth. Because proof verification requires only three elliptic curve pairings regardless of circuit size, the verifier workload remains constant and shows good promise to scale to large federations.

\textbf{Sampling soundness and detection guarantees.}
To verify that sampling $k = 32$ rows ensures statistical security against poisoning, we evaluate attestation across $19{,}992$ Monte Carlo draws ($v_{\max} = 1$, $u_{\max} = 8$; Table~\ref{tab:zk-full}). Honest batches pass with $98.51\%$ probability ($P(v \le 1)$), rarely triggering false rejections. Conversely, unconstrained attacks—such as channel roll ($21.29\%$ violations) and historical replay ($23.34\%$)—are rejected with $99.8\%$ certainty ($P(v > 1) = 0.998$), triggering Circom witness-generation aborts that prevent valid proof construction. While adaptive projection onto the invariant subspace lowers violations to $0.68\%$ to pass verification, this projection alone removes only part of the attack's poisoning damage. (Section~\ref{sec:results:removal}).

In summary, to answer RQ4: a participant can verify batch compliance in zero knowledge with practical overhead. By pairing Poseidon Merkle commitments with Groth16 row-sampling proofs ($k = 32$ of 1{,}024 rows), PA-FL Lite generates an 806-byte proof in 12.0\,s that verifies in 290\,ms. Honest batches pass with 98.5\% probability while unprojected attacks are rejected with 99.8\% certainty. Binding the model update to the committed batch remains future work.

\section{Discussion}\label{sec:discussion}
\subsection{Can Process Invariants Reliably Separate Honest from Fabricated Telemetry?}
In short, yes, honest batches pass the admission check and fabricated ones fail it. This \textit{physical separation} of batches succeeds because the fabrications preserve individual channel plausibility while breaking the algebraic relationships between coupled channels (Section~\ref{sec:results:rq1}). For instance, channel roll preserves every marginal distribution identically, and conservation scaling maintains each variable strictly within its nominal operational range; standard per-channel outlier filters or univariate distributional tests would admit both of these. Invariant admission rejects them because coupled states no longer agree---a running pump exhibits zero flow, or a fluctuating tank level contradicts the measured net flux. Conversely, honest telemetry, recorded under natural operational sensor noise, produces an empirical violation rate that peaks at $0.51\%$ (roughly half the 1\% gate), demonstrating that $\alpha$ must be calibrated against clean plant baselines: tightening $\alpha$ to $0.5\%$ triggers false rejections on honest client shards (Appendix~\ref{sec:app:alpha}).

This also establishes the benefit of data-space admission: internal physical consistency. Spliced sequences composed of authentic operational telemetry satisfy physical laws internally and pass the row-level check by construction, as do historical attack segments whose manipulated variables fall outside the instrumented invariant set (Section~\ref{sec:results:rq3}). Detecting such attacks requires cryptographic provenance rather than physical constraints. Furthermore, physical separation presupposes an instrumented, coupled industrial process. Under partial sensor coverage, some uncoordinated fabrications pass (such as conservation scaling on WADI, which lacks metered mass balances); in the absence of coupled actuator-hydraulic dynamics (HAI), the gate admits all batches and defence is in the hands of update-space aggregation. Crucially, facility operators can audit this boundary prior to deployment directly from the support and fit of invariants mined on their clean historical records.

\subsection{Why Are Update-Space Defences Blind to Poisoned Data?}
Intrusion detection for FL typically begins by checking the model updates produced by each individual plant for statistical anomalies. Updates that pass are admitted to the subsequent aggregation round. The primary issue here is that the update could have been produced by training on dishonest data; the update itself cannot discern either way. In other words, poisoning that enters through the training data cannot be picked up by rules that evaluate the update. In the next round the aggregation server will evaluate the update for potential anomalies and have none to find (Section~\ref{sec:results:rq2}).

It is well known that Byzantine-robust aggregation rules fail against clever adversaries~\cite{fang2020, baruch2019, shejwalkar2021} and our results deepen the failure. An adversary manipulating the data space bypasses most robust aggregation rules without needing to model or optimise against the server's defence. FLTrust is the exception, at the cost of a clean root dataset held by the server. The invariants play that role for the admission gate, as a compact physical reference rather than a dataset.

We have two main defences and argue that both are required because they operate on distinct objects: the admission gate evaluates the training batch before local computation, while the aggregation rule evaluates the parameter vectors afterward. Each defence is blind to the threat space governed by the other (Figure~\ref{fig:complementarity}). In this sense, the admission gate acts as a pre-admission filter, similar to how norm bounding is framed in secure aggregation~\cite{ma2025}. On the aggregation side, the federated learning literature typically defers poisoning to robust aggregation~\cite{hernandezramos2025, pecherle2026}. Our findings draw a line here: robust aggregation guards update-space manipulations but provides no barrier against data-space corruption.

\subsection{How Much Protection Does Process Physics Actually Provide?}\label{sec:disc:physics}
Process invariants began as hand-derived physical balances for testbed security to sound real-time alarms~\cite{adepu2016} and have since been refined through automated mining~\cite{feng2019, zhu2025}. The invariant check can be classified as rule-based by design as affine relations are deterministic, interpretable, and provable in zero knowledge. Admission is also an easier job than runtime detection: run offline once per round, it averages out transient row-level noise, a violation excludes an update (rather than shutting down the plant), and attackers must still fake coupled multivariate physics across actuators and flows~\cite{zhu2025}. Runtime monitoring and the gate are complementary: one protects the plant, the other the training.

Projection of the invariants removes about half of the damage (55\% pooled), while batch exclusion adds the rest on the wide set ($p = 0.008$;  Section~\ref{sec:results:removal}). As a result, the coverage ceiling that has traditionally constrained invariant-based detectors can be used as a tunable design parameter (Section~\ref{sec:results:narrow-wide}, Figure~\ref{fig:dial}). Operators can therefore audit and set the coverage boundary before training begins. When a batch is rejected, the cause is interpretable: the failed invariant is named, for example a pump reporting on while its flow meter reads zero. By the same token, protection extends only as far as the plant's metering.

So, how much protection does process physics actually provide? As much as plant instrumentation permits: the gate removes 66\%--92\% of targeted poisoning damage across the four main aggregation rules on coupled, metered balances, but offers no defence over unmonitored dynamics.

\subsection{Can We Verify the Data Without Seeing It?}
A fundamental challenge in designing a data-space admission gate is establishing trust: if a compromised client trains on poisoned telemetry, what mechanisms prevent it from falsely claiming that its data has passed the validation process?
The intuitive workaround, uploading telemetry to the coordinator so that the server can verify the invariants directly, is a privacy leak. Centralising operational SCADA logs violates participant privacy, and at least in the EU, violates critical-infrastructure cybersecurity directives and data-governance mandates~\cite{eu2022}. On the other hand, relying on an industrial client's honest updates is an open invitation to poisoning. Zero-knowledge attestation bridges the gap (Section~\ref{sec:results:rq4}): by evaluating process invariants inside an arithmetic circuit, a client can mathematically prove that its training batch respects the plant's governing physics without disclosing a sensor reading.

Other data-side defences do not remove this need, for example, a trusted execution environment attests a computation rather than the validity of its input data~\cite{chen2020}, and differential-privacy clipping acts on updates, which is norm clipping in its weak form, and among the rules we evaluate.

Early secure aggregation protocols explicitly left the validation of well-formed client inputs as an unresolved challenge~\cite{bonawitz2017}. Subsequent cryptographic input validation addressed this for update vectors by proving Euclidean norm bounds~\cite{lycklama2023, bell2023, ma2025}, while verifiable FL proved the integrity of server aggregation~\cite{wang2024, wang2025}. PA-FL Lite advances this work by moving the proved statement directly to the local training data, establishing batch telemetry as a verifiable object in federated systems. While proving full backpropagation training remains computationally prohibitive~\cite{peng2026}, proving batch compliance with physical predicates represents a compact, well-defined statement along that path, offering a practical intermediary without waiting for full proof-of-training to become tractable.

There is a trade-off to using ZK for physical admission: the proof that hides the data also hides the explanation trace. Invariant rejections on their own allow straightforward tracing to their root cause, but after the zk-SNARK, the operator only knows whether it is valid or invalid.
Revealing the underlying data would negate the ZK aspect and data privacy.

\subsection{Limitations}\label{sec:limitations}
The limitations of this study are multifaceted, reflecting the challenges of addressing both theoretical and practical considerations. Although we have sought to address existing gaps and maintain a positive mass flow, the experiments presented herein necessarily simplify the underlying complex and multidimensional systems.
With that said, a few words are due regarding adaptive adversaries, prototyping, and generalisability.

As in any setting with intelligence on both sides advancing offensive and defensive capabilities, adversaries can exploit blind spots: replayed sequences that obey the invariant set, regime-spliced data, and successfully projected batches all evade defences. The damage removal reported in Section~\ref{sec:results:removal} must therefore be read regarding covered attacks rather than universal protection. An adversary that directly optimises the poisoning objective within the feasible invariants and tailors updates to evade downstream aggregation rules, represents a stronger threat. It remains open how much poisoning capability survives when an adversary simultaneously optimises across both the physical invariants and the server's filters.

PA-FL Lite attests batch compliance over sampled rows but does not cryptographically bind the submitted update to the committed batch. A client committing honest telemetry while training on poisoned data would evade the gate, reducing to a naive attacker whose updates bypass most robust aggregation rules (Figure~\ref{fig:complementarity}). Proving the training step itself remains future work. Furthermore, the circuit checks a fixed-point representation of the invariants whose strict equivalence to floating-point constraints is argued and not proven~\cite{chen2024}. And while ZK conceals the raw telemetry batch, the public verification circuit embeds the invariant coefficients, disclosing aspects of plant topology to the coordinator. This leaves open whether the invariant equations themselves can be evaluated privately.

Lastly, data (and their public availability). The empirical evaluations rely on testbed facilities and simulated distribution networks governed by mass conservation and their publicly available datasets. Our federations have ten clients; because the gate is a per-client check its cost grows linearly with their number, and larger, heterogeneous federations are to be explored. Extending data-space admission to thermodynamic cycles, electrical grids with uncoupled dynamics (HAI), and live facilities remains future work.

\section{Conclusions}\label{sec:conclusion}
This work establishes that process physics can serve as an effective, privacy-preserving admission gate for federated intrusion detection in critical water infrastructure. Across physical SWaT and WADI records, automatically mined affine invariants admit all honest client shards while removing most of the targeted damage of historical attack replay, even when adversaries possess complete knowledge of the invariant set. By adjusting between narrow and wide invariant sets, operators explicitly control their defensive coverage boundaries prior to training. Finally, we show a zk-SNARK data-space admission check proves batch compliance in seconds on local clients and verifies in milliseconds at the coordinator's end, showing that data verification and privacy can be reconciled for sampled batches.


\subsection*{Author Contributions}
Conceptualisation, J.N., S.S., and B.O.; methodology, J.N.; software, J.N. and B.O.; validation, J.N., S.S., B.O., and S.T.; writing---original draft preparation, J.N. and S.S.; writing---review and editing, J.N., S.S., and S.T., All authors have read and agreed to the published version of the manuscript.

\subsection*{Funding}
This research received no external funding.

\subsection*{Data Availability}
The SWaT and WADI datasets analysed during this study are subject to third-party licensing restrictions and cannot be redistributed or republished. They are available for academic research upon formal request to the iTrust Centre for Research in Cyber Security at the Singapore University of Technology and Design (SUTD) at \url{https://itrust.sutd.edu.sg/itrust-labs_datasets/}. The BATADAL benchmark dataset is publicly accessible at \url{http://www.batadal.net/data.html}, and the HAI dataset is publicly available at \url{https://github.com/icsdataset/hai}. All source code, invariant mining scripts, zero-knowledge verification circuits, unit tests, and raw experimental evaluation results (JSON) supporting this study are publicly available in our GitHub repository at \url{https://github.com/RMIT-BDSL/pafl}.

\subsection*{Acknowledgements}
During the preparation of this manuscript, the authors used Claude Code (Fable 5.1) and Gemini (Flash 3.8) for assistance with coding and debugging, and figure generation. Writeful (Overleaf) was used for language editing and figure editing. The authors reviewed and edited the output and take full responsibility for the content of the published article.

\subsection*{Conflicts of Interest}
The authors declare no conflicts of interest.

\appendix

\section{Data Preparation and Verification Protocols}\label{sec:app:protocols}

This appendix records the channel handling applied to the two physical testbeds and specifies the two protocols by which the invariant sets are verified before any federated training (Section~\ref{sec:setup:protocols}).

\subsection{Testbed Channel Handling}
\label{sec:app:channels}

\textbf{SWaT analyser channels.} The Secure Water Treatment facility monitors its chemical dosing loops with Analyser Indicator Transmitters (AIT). Between the seven-day normal record and the attack record these channels drift by up to $18\sigma$ of their normal-record spread, a shift that reflects reagent depletion and probe fouling rather than any manipulation of the plant. Left in, the drift dominates an unsupervised detector's reconstruction error on the attack record: together with a numerical fault in the scaler (six constant actuators divided by a $10^{-8}$ spread), it held the clean detector at F1 0.27 with precision 0.16; excluding the AIT channels and flooring the scaler raised clean F1 to 0.756. The AIT channels are therefore excluded from detector training and from invariant mining.

\textbf{WADI channel handling.} Four channels of the WADI release are entirely empty in the historian export (\texttt{2\_LS\_001\_AL}, \texttt{2\_LS\_002\_AL}, \texttt{2\_P\_001\_STATUS}, \texttt{2\_P\_002\_STATUS}) and are dropped; a few analyser channels have short gaps, which are filled from the neighbouring rows.

\subsection{Separation Protocol}
\label{sec:app:separation}

The separation protocol asks whether the mined invariants distinguish honest telemetry from fabricated telemetry (without rejecting honest batches):
\begin{enumerate}
  \item \textbf{Batch sampling.} For each record (SWaT, WADI, BATADAL and HAI), 200 batches of $N = 1{,}000$ consecutive rows are drawn at random start positions from the attack-free record.
 \item \textbf{Fabrications.} Each batch is fabricated in the four ways defined in Section~\ref{sec:method:attacks}: channel roll by 60 rows (5\,min at the 5\,s stride; 7 rows on BATADAL and HAI), regime permutation over four $k$-means regimes, conservation scaling of the flows by 1.3, and regime splicing into twelve reordered segments.
  \item \textbf{Outcome.} Every honest and fabricated batch is scored against the calibrated invariant set under the 1\% admission rule. The protocol records the fraction of violating rows per batch, the honest false-rejection rate and the rejection rate of each fabrication (Figure~\ref{fig:c1-separation}).
\end{enumerate}

\subsection{Attack Coverage Audit}
\label{sec:app:coverage}

The coverage audit determines which of the plant's own labelled attacks an invariant set can see:
\begin{enumerate}
  \item \textbf{Segmentation.} Each contiguous run of labelled attack rows at the 5\,s stride is one attack: 35 on SWaT and 14 on WADI (Section~\ref{sec:setup:protocols} gives the counting convention).
  \item \textbf{Scoring.} Every attack is scored under each invariant set. An attack counts as \emph{covered} when more than 1\% of its rows violate at least one invariant, the same rule the gate applies to a batch.
  \item \textbf{Honest rate.} The honest violation rate of each set is measured on the first 60{,}000 rows after the invariant slice of the normal record; it is the cost side of the coverage dial (Table~\ref{tab:coverage}).
\end{enumerate}

\section{Targeted-Recall Damage and Removal and Detector Performance per Rule}\label{sec:app:targeted-removal}
Table~\ref{tab:removal} shows the targeted-recall damage and removal across aggregation rules. For each aggregation rule, attacker, and invariant set, the damage of the naive attack in the targeted recall points, the share of that damage removed by projection alone and by the deployed gate, and the share of malicious updates the rule admitted in each federation. The removal ratios are damage-weighted over the informative seeds.

\begin{table}[H]
\centering
\small
\caption{Targeted-recall damage and its removal per aggregation rule on SWaT (20 seeds). $n$: informative seeds (naive damage $\geq 1$ point); damage is their mean, in points. Rule accepts: share of the attacker's updates the rule aggregates. Removal: share of naive damage removed, with 95\% bootstrap intervals; $^{*}p<0.05$, $^{**}p<0.01$, $^{***}p<0.001$ (exact one-sided Wilcoxon test, gated vs naive). Values above 100\% lie within seed variance. $^\dagger$Krum accepts no malicious update, so its damage is not poisoning.}
\label{tab:removal}
\setlength{\tabcolsep}{3.5pt}
\footnotesize
\setlength{\tabcolsep}{3pt}
\begin{tabular}{@{}l c c r r r@{\hspace{3pt}[}r@{,\hspace{2pt}}r@{]}l@{}}
\toprule
 & & & & \multicolumn{4}{c}{Damage removed (\%)} & \\
\cmidrule(lr){5-8}
Rule & \begin{tabular}[b]{@{}c@{}}Rule\\{}accepts\end{tabular} & $n$ & \begin{tabular}[b]{@{}c@{}}Naive\\{}damage\end{tabular} & Proj. & \multicolumn{3}{c}{Gate [95\% CI]} & \\
\midrule
\multicolumn{9}{@{}l}{\emph{Replay, wide set}} \\
\quad FedAvg & 1.00 & 11 & 11.7 & 59 & 82 & 61 & 102 & $^{***}$ \\
\quad Median & 1.00 & 13 & 6.2 & 50 & 68 & 41 & 99 & $^{***}$ \\
\quad Norm clip & 1.00 & 11 & 8.2 & 75 & 92 & 74 & 103 & $^{***}$ \\
\quad Trimmed mean & 0.34 & 13 & 7.0 & 59 & 77 & 56 & 98 & $^{***}$ \\
\quad FLTrust & 0.22 & 5 & 2.2 & 47 & 63 & 34 & 100 & $^{*}$ \\
\quad FoolsGold & 0.66 & 8 & 10.6 & $-$14 & 5 & $-$53 & 53 &  \\
\quad Krum$^{\dagger}$ & 0.00 & 3 & 8.4 & 0 & 29 & 0 & 100 &  \\
\addlinespace
\multicolumn{9}{@{}l}{\emph{Channel roll, wide set}} \\
\quad FedAvg & 1.00 & 10 & 9.1 & 40 & 82 & 57 & 105 & $^{**}$ \\
\quad Median & 1.00 & 10 & 6.7 & 53 & 71 & 39 & 100 & $^{***}$ \\
\quad Norm clip & 1.00 & 7 & 8.5 & 66 & 88 & 61 & 100 & $^{*}$ \\
\quad Trimmed mean & 0.34 & 11 & 6.1 & 43 & 66 & 34 & 96 & $^{**}$ \\
\quad FLTrust & 0.15 & 5 & 2.3 & 11 & 34 & $-$2 & 81 &  \\
\quad FoolsGold & 0.90 & 9 & 11.1 & $-$16 & 21 & $-$43 & 73 &  \\
\quad Krum$^{\dagger}$ & 0.00 & 3 & 8.4 & 0 & 29 & 0 & 100 &  \\
\addlinespace
\multicolumn{9}{@{}l}{\emph{Replay, narrow set}} \\
\quad FedAvg & 1.00 & 11 & 11.7 & 44 & 53 & 30 & 77 & $^{**}$ \\
\quad Median & 1.00 & 13 & 6.2 & 49 & 61 & 29 & 86 & $^{***}$ \\
\quad Norm clip & 1.00 & 11 & 8.2 & 67 & 70 & 37 & 97 & $^{***}$ \\
\quad Trimmed mean & 0.34 & 13 & 7.0 & 45 & 57 & 33 & 81 & $^{***}$ \\
\quad FLTrust & 0.22 & 5 & 2.2 & 35 & 35 & 5 & 77 &  \\
\quad FoolsGold & 0.66 & 8 & 10.6 & $-$29 & $-$16 & $-$82 & 46 &  \\
\quad Krum$^{\dagger}$ & 0.00 & 3 & 8.4 & 0 & 29 & 0 & 100 &  \\
\addlinespace
\multicolumn{9}{@{}l}{\emph{Channel roll, narrow set}} \\
\quad FedAvg & 1.00 & 10 & 9.1 & 35 & 47 & 21 & 78 & $^{**}$ \\
\quad Median & 1.00 & 10 & 6.7 & 45 & 58 & 27 & 85 & $^{**}$ \\
\quad Norm clip & 1.00 & 7 & 8.5 & 58 & 63 & 27 & 91 & $^{*}$ \\
\quad Trimmed mean & 0.34 & 11 & 6.1 & 56 & 71 & 42 & 92 & $^{**}$ \\
\quad FLTrust & 0.15 & 5 & 2.3 & $-$14 & $-$14 & $-$44 & 19 &  \\
\quad FoolsGold & 0.90 & 9 & 11.1 & $-$23 & $-$5 & $-$76 & 36 &  \\
\quad Krum$^{\dagger}$ & 0.00 & 3 & 8.4 & 0 & 29 & 0 & 100 &  \\
\bottomrule
\end{tabular}

\end{table}

Table~\ref{tab:detector} reports the global detector's own metrics in every federation (FedAvg). Poisoning barely moves F1 or AUC-PR but lowers recall on the targeted attacks, which is why removal is measured on targeted recall; gated federations match the honest-only reference, so the gate costs no detection utility.

\begin{table}[H]
\centering
\small
\caption{Detector metrics of the global model per federation (FedAvg; mean $\pm$ standard deviation over seeds). The alarm threshold is calibrated on clean data; AUC-PR is threshold-free. The clean and honest-only federations do not involve the attacker; BATADAL has no honest-only federation and no untargeted attack windows.}
\label{tab:detector}
\setlength{\tabcolsep}{6.5pt}
\begin{tabular}{@{}l c c c c@{}}
\toprule
Federation & F1 & AUC-PR & \begin{tabular}[b]{@{}c@{}}Targeted\\{}recall\end{tabular} & \begin{tabular}[b]{@{}c@{}}Untargeted\\{}recall\end{tabular} \\
\midrule
\multicolumn{5}{@{}l}{\emph{SWaT, wide set} (20 seeds)} \\
\quad clean & 0.771 $\pm$ 0.010 & 0.797 $\pm$ 0.005 & 0.206 $\pm$ 0.112 & 0.719 $\pm$ 0.014 \\
\quad honest-only & 0.771 $\pm$ 0.016 & 0.797 $\pm$ 0.005 & 0.198 $\pm$ 0.119 & 0.715 $\pm$ 0.018 \\
\quad replay, naive & 0.773 $\pm$ 0.008 & 0.780 $\pm$ 0.004 & 0.133 $\pm$ 0.094 & 0.711 $\pm$ 0.015 \\
\quad replay, physics-aware & 0.770 $\pm$ 0.010 & 0.790 $\pm$ 0.012 & 0.171 $\pm$ 0.108 & 0.705 $\pm$ 0.014 \\
\quad replay, gated & 0.770 $\pm$ 0.014 & 0.792 $\pm$ 0.011 & 0.186 $\pm$ 0.118 & 0.711 $\pm$ 0.017 \\
\quad roll, naive & 0.774 $\pm$ 0.007 & 0.782 $\pm$ 0.005 & 0.152 $\pm$ 0.102 & 0.710 $\pm$ 0.015 \\
\quad roll, physics-aware & 0.771 $\pm$ 0.011 & 0.797 $\pm$ 0.006 & 0.173 $\pm$ 0.105 & 0.707 $\pm$ 0.014 \\
\quad roll, gated & 0.771 $\pm$ 0.014 & 0.797 $\pm$ 0.006 & 0.192 $\pm$ 0.118 & 0.712 $\pm$ 0.016 \\
\addlinespace
\multicolumn{5}{@{}l}{\emph{WADI} (3 seeds)} \\
\quad clean & 0.260 $\pm$ 0.002 & 0.352 $\pm$ 0.008 & 0.197 $\pm$ 0.091 & 0.096 $\pm$ 0.142 \\
\quad honest-only & 0.264 $\pm$ 0.006 & 0.348 $\pm$ 0.011 & 0.198 $\pm$ 0.091 & 0.104 $\pm$ 0.135 \\
\quad replay, naive & 0.256 $\pm$ 0.002 & 0.305 $\pm$ 0.014 & 0.192 $\pm$ 0.093 & 0.097 $\pm$ 0.144 \\
\quad replay, physics-aware & 0.256 $\pm$ 0.001 & 0.304 $\pm$ 0.012 & 0.192 $\pm$ 0.093 & 0.097 $\pm$ 0.143 \\
\quad replay, gated & 0.262 $\pm$ 0.008 & 0.332 $\pm$ 0.032 & 0.196 $\pm$ 0.089 & 0.104 $\pm$ 0.135 \\
\quad roll, naive & 0.258 $\pm$ 0.003 & 0.309 $\pm$ 0.016 & 0.194 $\pm$ 0.092 & 0.098 $\pm$ 0.141 \\
\quad roll, physics-aware & 0.257 $\pm$ 0.002 & 0.312 $\pm$ 0.019 & 0.194 $\pm$ 0.093 & 0.097 $\pm$ 0.142 \\
\quad roll, gated & 0.262 $\pm$ 0.007 & 0.333 $\pm$ 0.031 & 0.197 $\pm$ 0.090 & 0.103 $\pm$ 0.136 \\
\addlinespace
\multicolumn{5}{@{}l}{\emph{BATADAL} (3 seeds)} \\
\quad clean & 0.470 $\pm$ 0.000 & 0.653 $\pm$ 0.005 & 0.337 $\pm$ 0.000 & -- \\
\quad replay, naive & 0.458 $\pm$ 0.019 & 0.619 $\pm$ 0.011 & 0.323 $\pm$ 0.024 & -- \\
\quad replay, physics-aware & 0.458 $\pm$ 0.019 & 0.619 $\pm$ 0.010 & 0.323 $\pm$ 0.024 & -- \\
\quad replay, gated & 0.458 $\pm$ 0.019 & 0.619 $\pm$ 0.010 & 0.323 $\pm$ 0.024 & -- \\
\quad roll, naive & 0.433 $\pm$ 0.036 & 0.593 $\pm$ 0.014 & 0.306 $\pm$ 0.032 & -- \\
\quad roll, physics-aware & 0.443 $\pm$ 0.014 & 0.595 $\pm$ 0.008 & 0.314 $\pm$ 0.011 & -- \\
\quad roll, gated & 0.443 $\pm$ 0.014 & 0.595 $\pm$ 0.008 & 0.314 $\pm$ 0.011 & -- \\
\bottomrule
\end{tabular}

\end{table}

\section{Sensitivity to the Admission Threshold}\label{sec:app:alpha}
\begin{figure}[htb]
\centering
\includegraphics[width=\textwidth]{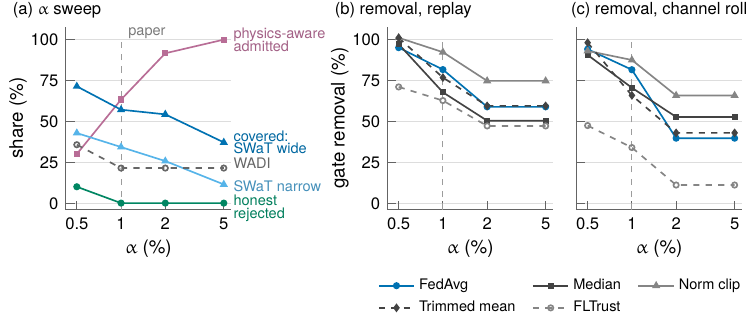}
\caption{Sensitivity to the admission threshold $\alpha$ (SWaT wide set unless stated; 20 seeds). (a)~Distinct honest shards rejected, physics-aware batches admitted, and labelled attacks covered. No honest shard is rejected on SWaT narrow, WADI or BATADAL at any $\alpha$; at 0.5\% one SWaT wide-set shard (0.51\% violating rows) is rejected. (b,~c)~Share of naive damage removed by the gate, per rule; at $\alpha \geq 2\%$ the gate excludes no attacker in an informative seed, so removal equals the projection-only value. The dashed line marks our paper's $\alpha = 1\%$.}
\label{fig:alpha}
\end{figure}

An alpha sweep is shown in Figure~\ref{fig:alpha} repeating the gated federations at $\alpha = 0.5$, 2 and 5\% (SWaT wide set, 20 seeds). Tightening $\alpha$ to 0.5\% rejects one honest shard while loosening it to 2\% or more admits nearly every projected batch. The paper's $\alpha = 1\%$ is the tightest setting tested that rejects no honest shard.

\section{Zero-Knowledge Circuit Arithmetisation (PA-FL Lite)}\label{sec:app:zk}

This appendix gives the protocol sequence and the arithmetisation of the PA-FL Lite Circom circuit, including fixed-point encoding, the biased representation that prevents field wrap-around, the Poseidon leaf hashing, and the constraint breakdown and prover cost in Table~\ref{tab:zk-full}.

\begin{figure}[htb]
\centering
\includegraphics[width=\textwidth]{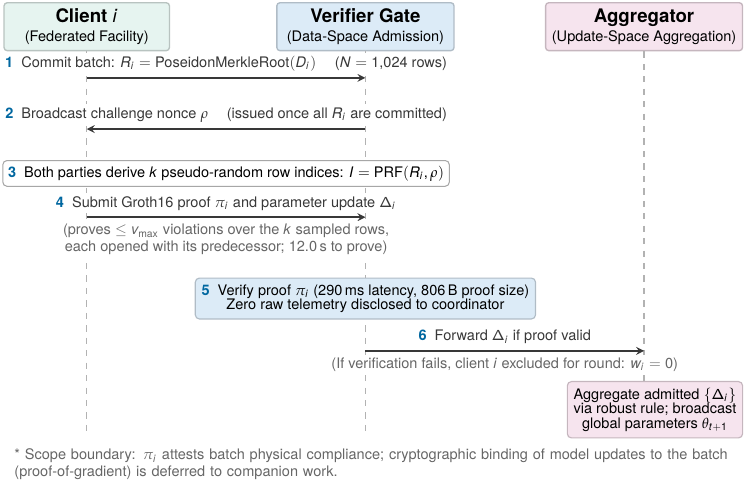}
\caption{PA-FL Lite protocol round. The client commits to the training batch ($N = 1{,}024$ rows) as a Poseidon Merkle root; once every commitment is in, the verifier broadcasts a challenge nonce, and both parties derive $k$ pseudo-random row indices. A Groth16 proof is returned that at most $v_{\max}$ sampled rows violate an invariant, and that at most $u_{\max}$ checks are inapplicable. The verifier checks the proof and the aggregator combines the admitted updates with a robust rule; a client whose proof fails is excluded for the round ($w_i = 0$). Timings are for $k = 32$ (Table~\ref{tab:zk-headline}).}
\label{fig:protocol}
\end{figure}

\subsection{Circuit Pipeline and R1CS Optimisations}
To validate telemetry inside a rank-1 constraint system within an edge device's budget, the circuit uses four arithmetisation choices:
\begin{enumerate}
  \item \textbf{Constants baked into the template.} The invariant parameters, the coefficient matrices $(J_{\mathrm{prev}}, J_{\mathrm{cur}})$, offsets $c_j$ and tolerances $\varepsilon_j$, are compiled into the Circom template as numerical constants. The affine residuals $r_j = J_{\mathrm{prev},j}\,x[i-1] + J_{\mathrm{cur},j}\,x[i] + c_j$ are then linear combinations of witness values and cost no multiplication gates. The public inputs are only $(R_i, I, v_{\max}, u_{\max})$, $k + 3$ field elements (35 at $k = 32$).
  \item \textbf{Fixed-point scaling with a biased encoding.} Sensor readings and tolerances are scaled to integers by $S = 2^{16}$. Field elements of $\mathbb{F}_p$ ($p \approx 2^{254}$ on the BN128 scalar field) have no sign, so a negative value would wrap to near $p$; and projected batches do carry negative values, since the adversary's projection drives unconstrained flow channels below zero (in one projected SWaT batch, FIT401 reads $-62\,\mathrm{m^3/h}$). Every value therefore enters the circuit as $\hat{x} = \lfloor x \cdot 2^{16} \rceil + 2^{31}$ and is range-checked to 32 bits, and every residual stays below $2^{53} \ll p$, so field arithmetic reproduces signed integer arithmetic exactly.
  \item \textbf{Chunked Poseidon leaf hashing.} A Poseidon permutation takes at most 16 inputs, so the 18 constrained channels of a row are hashed in two chunks whose digests are combined with a precomputed digest of the row's unconstrained channels. A leaf costs 1{,}095 constraints, against about 1{,}500 for a single wide sponge.
  \item \textbf{Soft violation accumulation.} Instead of asserting $|r_j| \le \varepsilon_j$ row by row, which would make the circuit unsatisfiable at the first violation, each check is a 56-bit comparison on the biased residual that emits a Boolean flag. The flags, and the flags of inapplicable actuator checks, are summed over the $k$ samples and compared with the public budgets $v_{\max}$ and $u_{\max}$ once, at the circuit boundary.
\end{enumerate}

\subsection{Constraint Breakdown and Prover Cost}
At $k = 32$ samples over a batch of $N = 1{,}024$ rows (Merkle depth 10), the circuit has 297{,}736 R1CS constraints, 9{,}304 per sample. The two Merkle path openings ($2 \times 2{,}420$) and the two leaf hashes ($2 \times 1{,}095$) take 7{,}030 of these, about 76\%; the 36 range checks on the two sampled rows take 1{,}152 (12\%), the nine tolerance checks 1{,}035 (11\%), and the applicability and index-bit checks the remaining 80. Benchmarked on an Apple M3 Pro (Circom~2.1.9, SnarkJS~0.7.4, Groth16 over BN128, Powers of Tau $2^{20}$), proving takes 12.0\,s at 3.8\,GB peak memory; verification is 290\,ms independent of $k$; and the Groth16 proof consists of three group elements ($A \in G_1$, $B \in G_2$, $C \in G_1$), consuming 256 bytes in binary and 806 bytes in SnarkJS's JSON encoding. At $k = 16$, the circuit has 148{,}888 constraints and proves in 6.55\,s with the same verification cost (Table~\ref{tab:zk-full}).

\begin{table}[H]
\centering
\small
\caption{PA-FL Lite full benchmark across sample sizes $k$. Attestation outcomes are for the seed-0 SWaT batches under a budget of $v_{\max} = 1$ violations and $u_{\max} = 8$ inapplicable checks; sampling soundness is estimated from 19{,}992 random index draws over 56 honest windows.}
\label{tab:zk-full}
\setlength{\tabcolsep}{4pt}
\begin{tabular}{@{}l r r@{}}
\toprule
\textbf{Protocol Metric / Circuit Property} & \textbf{$k = 16$ samples} & \textbf{$k = 32$ samples} \\
\midrule
R1CS arithmetic constraints (total) & 148{,}888 & 297{,}736 \\
Constraints per sample ($2$ paths, $2$ leaves, physics) & 9{,}306 & 9{,}304 \\
Public / private circuit inputs & 19 / 928 & 35 / 1{,}856 \\
Circuit compilation time (Circom) & 7.47\,s & 15.0\,s \\
Proving key size (\texttt{.zkey}) & 108.7\,MB & 217.4\,MB \\
Verification key size (\texttt{.vk}) & 6.1\,kB & 8.9\,kB \\
Witness generation (honest batch) & 0.41\,s & 0.56\,s \\
Proof generation time $\pi_i$ (peak RAM) & 6.55\,s (2.3\,GB) & 12.0\,s (3.8\,GB) \\
Proof verification time (coordinator) & 300\,ms & 290\,ms \\
Proof size (Groth16 $\pi_i$) & 808\,B & 806\,B \\
\midrule
\multicolumn{3}{@{}l}{\emph{Empirical Attestation Outcome ($v_{\max} = 1, u_{\max} = 8$ budget)}} \\
Honest batch ($0.39\%$ violating rows) & Verified ($0 / 0$) & Verified ($0 / 0$) \\
Channel roll ($21.29\%$ violating rows) & Hard-abort ($6 / 1$) & Hard-abort ($18 / 1$) \\
Historical replay ($23.34\%$ violating rows) & Hard-abort ($5 / 0$) & Hard-abort ($6 / 0$) \\
Physics-aware projected ($0.68\%$ violating rows) & Verified ($0 / 0$) & Verified ($0 / 0$) \\
\midrule
\multicolumn{3}{@{}l}{\emph{Monte Carlo Sampling Soundness ($19{,}992$ draws)}} \\
Honest batch acceptance ($P(v \le 1)$) & 99.41\,\% & 98.51\,\% \\
Detection of channel roll ($P(v > 1)$) & 93.5\,\% & 99.8\,\% \\
Detection of historical replay ($P(v > 1)$) & 91.7\,\% & 99.8\,\% \\
\bottomrule
\end{tabular}

\end{table}

\subsection{Binding the Update to the Committed Batch}\label{sec:app:binding}
PA-FL Lite proves a statement about a committed batch, while the aggregator receives an update, and nothing in the protocol ties the two: a client can commit an honest batch, pass verification, and submit an update trained on other data. Such a client is equivalent to the naive attacker (Section~\ref{sec:method:attacks}), and the gate does not stop it. Narrowing this gap is an important direction for future work. Two promising pathways stand out: zero-knowledge proofs of training~\cite{peng2026}, which cryptographically bind the update to the committed batch but currently incur substantial circuit overheads for backpropagation, and probabilistic spot-auditing, where selected rounds are verified by re-executing training on revealed batches. In the interim, robust aggregation remains the active backstop: in our empirical sweep, FLTrust limited data poisoning damage to 1.4--2.0 percentage points of targeted recall, whereas FedAvg, coordinate median, and norm clipping admitted every poisoned update (Figure~\ref{fig:complementarity}).

\clearpage
\bibliographystyle{plainnat}
\bibliography{pafl-refs}

\end{document}